\documentclass[%
 reprint,
 superscriptaddress,
 amsmath,
 aps,
 prb,
 longbibliography
]{revtex4-2}

\usepackage{graphicx, color} 
\usepackage{newtxtext, newtxmath}
\usepackage{bm} 
\usepackage{braket, mathtools}
\usepackage{multirow}
\usepackage[
 colorlinks=true,
 citecolor=blue,
 linkcolor=blue
]{hyperref}
\usepackage[export]{adjustbox} 
\usepackage{cellspace} 

\graphicspath{{./images/}}

\DeclareMathOperator{\sgn}{sgn}
\DeclareMathOperator{\tr}{tr}

\newcommand{\ii}{\mathrm{i}}
\newcommand{\ee}{\mathrm{e}}
\let\tilde\widetilde

\begin{document}

\title{%
 Phase-modulated superconductivity via altermagnetism
}

\author{Shuntaro Sumita}
\email[]{s-sumita@g.ecc.u-tokyo.ac.jp}
\affiliation{%
 Department of Basic Science, The University of Tokyo, Meguro, Tokyo 153-8902, Japan
}%
\affiliation{%
 Komaba Institute for Science, The University of Tokyo, Meguro, Tokyo 153-8902, Japan
}%
\affiliation{%
 RIKEN Center for Emergent Matter Science, Wako, Saitama 351-0198, Japan
}%

\author{Makoto Naka}
\affiliation{%
 School of Science and Engineering, Tokyo Denki University, Ishizaka, Saitama 350-0394, Japan
}%

\author{Hitoshi Seo}
\affiliation{%
 RIKEN Center for Emergent Matter Science, Wako, Saitama 351-0198, Japan
}%

\date{\today}

\begin{abstract}
 Stimulated by recent interest in altermagnets, a novel class of antiferromagnets with macroscopic time-reversal symmetry breaking, we investigate the coexistence of altermagnetism and superconductivity.
 By developing a Ginzburg--Landau theory based on microscopic models, we show that a \textit{phase}-modulated Fulde--Ferrell superconducting state is stabilized via altermagnetic spin splitting, in contrast to the typical \textit{amplitude}-modulated states that occur under the uniform Zeeman field.
 We apply our framework to different models to compare the resulting phase diagrams: a two-sublattice model with altermagnetic order, a continuum model with an anisotropic Zeeman field mimicking altermagnetic spin splitting, and a conventional square-lattice model with two kinds of anisotropic Zeeman fields.
 We show that the multisublattice structure is crucial for realizing the phase-modulated superconductivity, and highlight spin-split altermagnets as a promising platform for exploring this exotic superconductivity without external magnetic fields.
\end{abstract}

\maketitle

\section{Introduction}
Theoretical findings of characteristic properties in collinear antiferromagnets such as spin-split dispersion relation in electronic~\cite{Noda2016, Okugawa2018, Ahn2019, Naka2019, Hayami2019} and magnon~\cite{Naka2019} bands leading to spin current generation~\cite{Ahn2019, Naka2019}, and the anomalous Hall effect~\cite{Solovyev1997, Smejkal2020, Naka2020} led to the concept of `'`altermagnetism''~\cite{Smejkal2022_Sep, Smejkal2022_Dec}, which is now attracting significant interest.
Altermagnets exhibit macroscopic time-reversal symmetry breaking below the N\'{e}el temperature because the distinct sublattices hosting up and down spins are connected not by inversion or any translation operation but by other space group operations (such as rotation, screw, or glide).
A wide variety of materials is considered as their candidates: rutiles~\cite{Noda2016, Ahn2019, Smejkal2020}, organic antiferromagnets $\kappa$-(ET)$_2X$~\cite{Naka2019, Naka2020, Seo2021}, perovskites~\cite{Solovyev1997, Okugawa2018, Naka2021, Naka2025_review}, and MnTe~\cite{Hariki2024, Gonzalez_Betancourt2023, Krempask2024, Osumi2024, Liu2024}.
Systems showing magnetic octupole order~\cite{Suzuki2017, Suzuki2019, Hayami2018, HikaruWatanabe2018} are also relevant to the study of altermagnets.

One of the intriguing subjects is the coexistence of altermagnetism and superconductivity.
Indeed, recent theoretical studies have proposed that an altermagnetic order can induce Fulde--Ferrell--Larkin--Ovchinnikov (FFLO) superconductivity~\cite{FF, LO} or pair-density-wave (PDW) superconductivity, where the Cooper pairs possess finite center-of-mass (COM) momentum~\cite{Sumita2023, Zhang2024, Chakraborty2024, Chakraborty2024_arXiv_1, Chakraborty2024_arXiv_2, Sim2024, Hong2025, Mukasa2025, Iorsh2025, Hu2025_arXiv_1, Hu2025_arXiv_2, Ouassou2023, Giil2024, Banerjee2024}, stabilized by the altermagnetic spin splitting.
Such an FFLO state in the presence of anisotropic spin splitting was proposed in the context of a spin-nematic order about a decade ago~\cite{Soto-Garrido2014, Gukelberger2014}.
A compelling point about these exotic states is that they can be realized without an external magnetic field and therefore are free from vortices, in clear contrast with the original proposal of FFLO superconductivity under external magnetic field.
Here we note that the altermagnetic spin ordering, by definition as mentioned above, requires at least two sublattice degrees of freedom.
Without such an ingredient, however, an anisotropic spin splitting can be introduced by a momentum-dependent Zeeman field, which is somewhat artificial and sometimes confusedly used in recent literature. 

An issue in the FFLO state in altermagnets is its spatial structure, which is still controversial.
In general, there are various possibilities for spatially modulated superconductivity: a phase-modulated FF state [with spatially varied order parameter $\eta(\bm{R}) \sim \ee^{2\ii\bm{Q} \cdot \bm{R}}$], an amplitude-modulated LO state [$\eta(\bm{R}) \sim \cos(2\bm{Q} \cdot \bm{R})$], a bidirectional PDW state [$\eta(\bm{R}) \sim \cos(2\bm{Q} \cdot \bm{R}) + \cos(2\bm{Q}' \cdot \bm{R})$], etc.
On the basis of the Ginzburg--Landau (GL) theory, Ref.~\cite{Soto-Garrido2014} showed that the LO or bidirectional PDW state is stabilized under an anisotropic spin splitting, while Ref.~\cite{Sim2024} proposed the FF state in altermagnets.
In both works, the anisotropic spin splitting is introduced by a momentum-dependent Zeeman field in the model, and the sublattice degree of freedom is not taken into account.

To address this issue, in this paper we microscopically develop the GL theory based on a \textit{multisublattice} tight-binding model.
As mentioned above, most of the previous studies have overlooked such a sublattice degree of freedom but instead introduced a momentum-dependent Zeeman field to describe the anisotropic spin splitting in altermagnets~\cite{Zhang2024, Chakraborty2024, Chakraborty2024_arXiv_2, Sim2024, Hong2025, Mukasa2025, Iorsh2025, Hu2025_arXiv_1, Hu2025_arXiv_2}.
Here we reveal that the phase-modulated FF superconductivity, rather than amplitude-modulated states, is stabilized in the presence of altermagnetic order.
This is a distinctive characteristic of altermagnets, because typically the amplitude-modulated state is more stable in most superconductors under the uniform Zeeman field~\cite{Shimahara1998, Matsuda_Shimahara_review, Buzdin1997, Agterberg2001}.
Furthermore, we demonstrate that the multisublattice degree of freedom is actually essential for realizing the FF state, by comparing our multisublattice model with other models introducing the anisotropic Zeeman field as in the literature.

The rest of this paper is organized as follows.
In Sec.~\ref{sec:model}, we introduce three kinds of two-dimensional (2D) models studied here.
The main target is a tetragonal model with two sublattices that is one of the minimal models for altermagnets (Sec.~\ref{sec:model_square2}).
Because of asymmetric hoppings in the two sublattices with different directions, the energy band has asymmetric spin splitting when an antiferromagnetic molecular field is switched on.
To uncover the effect of the multisublattice property, we then introduce a continuum model (Sec.~\ref{sec:model_continuum}) and a conventional square lattice model (Sec.~\ref{sec:model_square1}).
We incorporate the momentum-dependent Zeeman field by hand into both models, which lack the sublattice degree of freedom.
Particularly in the conventional square lattice model, we compare the cases with two forms of momentum dependences which we call $1\bm{k}$- and $2\bm{k}$-$d$-wave Zeeman fields, the latter of which properly corresponds to the structure of anisotropic spin splitting in the two-sublattice tetragonal model.
Next, on the basis of the mean-field (MF) approximation to introduce superconductivity, we microscopically derive the GL theory and compute the coefficients in the GL free energy for each model (Sec.~\ref{sec:GL}).
The models, their spin split structures, and the resulting superconducting states are summarized in Table~\ref{tab:models}: our analyses show that, as the molecular field or the Zeeman field is increased, the phase-modulated FF state becomes stable in the two-sublattice tetragonal model [Table~\ref{tab:models}(a)] and the conventional square lattice model in a $2\bm{k}$-$d$-wave Zeeman field [(c2)], whereas the amplitude-modulated state is stabilized in the other cases [(b) and (c1)].
These results highlight the crucial role of the multisublattice degree of freedom.
A summary and discussion are given in Sec.~\ref{sec:summary}.

\begin{table*}
 \caption{Characteristics of (a) two-sublattice tetragonal, (b) continuum, and (c) conventional square lattice models considered in this paper. In the second column, we show illustrations of the square lattices for (a) and (c), where the unit cells are represented by green squares, and the energy dispersion for (b). In the third column, we refer to molecular (Zeeman) fields for anisotropic spin splitting, which depend on the momentum for (b) and (c) but not for (a). In particular, we treat (c1) $1\bm{k}$-$d$-wave and (c2) $2\bm{k}$-$d$-wave Zeeman fields in the conventional square lattice model. Contour maps of the spin splitting subtracting the down-spin energy from the up-spin energy are shown in the fourth column. In the fifth column, the resultant spatial structures of nonuniform superconductivity in a large field from our GL theory are summarized.}
 \label{tab:models}
 \setlength\cellspacetoplimit{1mm}
 \setlength\cellspacebottomlimit{1mm}
 \begin{tabular}{cSccScc} \hline\hline
  Model & Transfer integral (dispersion) & Anisotropic field $\hat{H}_{\mathrm{mag}}$ & Spin splitting & $\bm{Q} \neq \bm{0}$ superconductivity \\ \hline
  \parbox{26mm}{(a) two-sublattice tetragonal}
  & \includegraphics[valign=c, width=32mm]{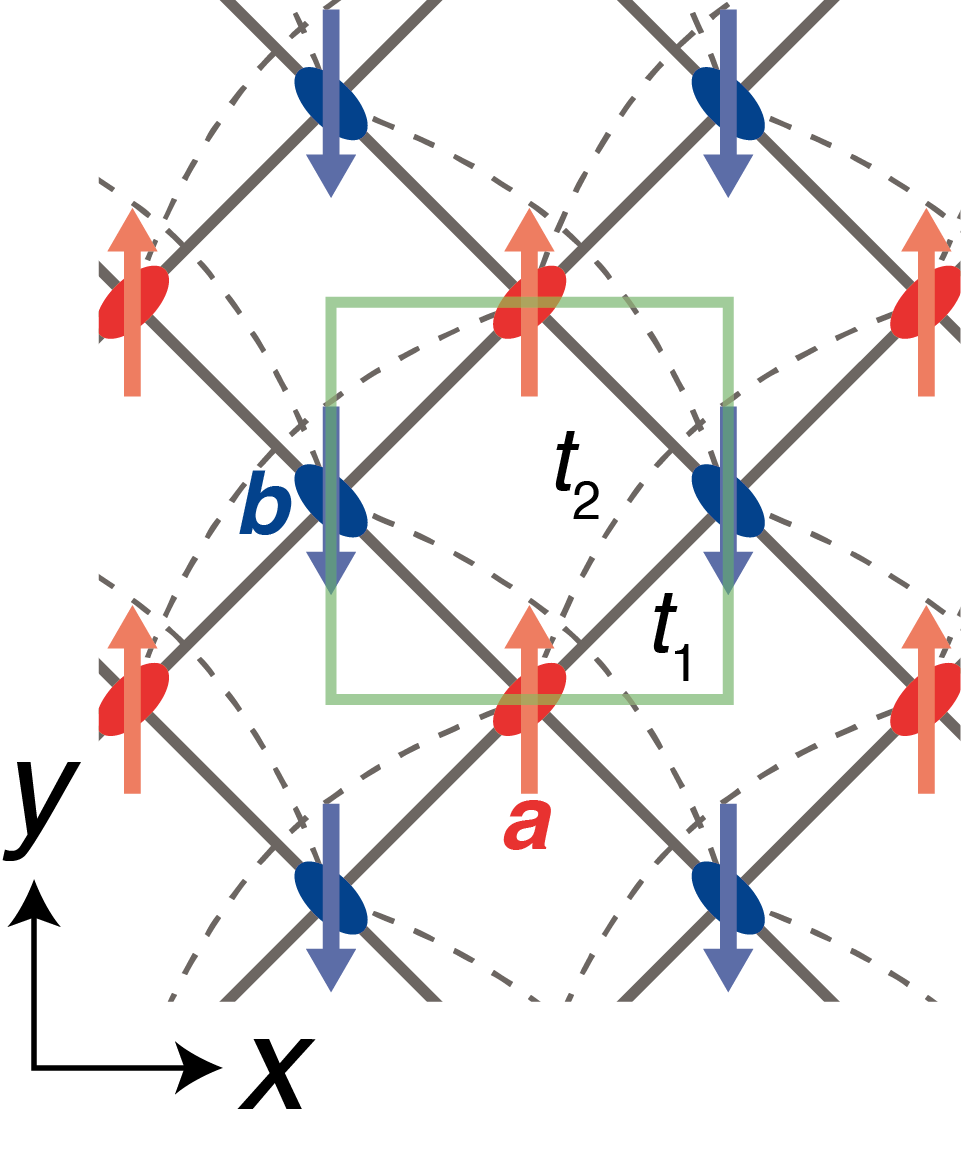}
  & \parbox{40mm}{$- h \hat{\tau}_z \otimes \hat{\sigma}_z$ \\ {[Eq.~\eqref{eq:Hamiltonian_square2_mag}]}}
  & \includegraphics[valign=c, width=28mm]{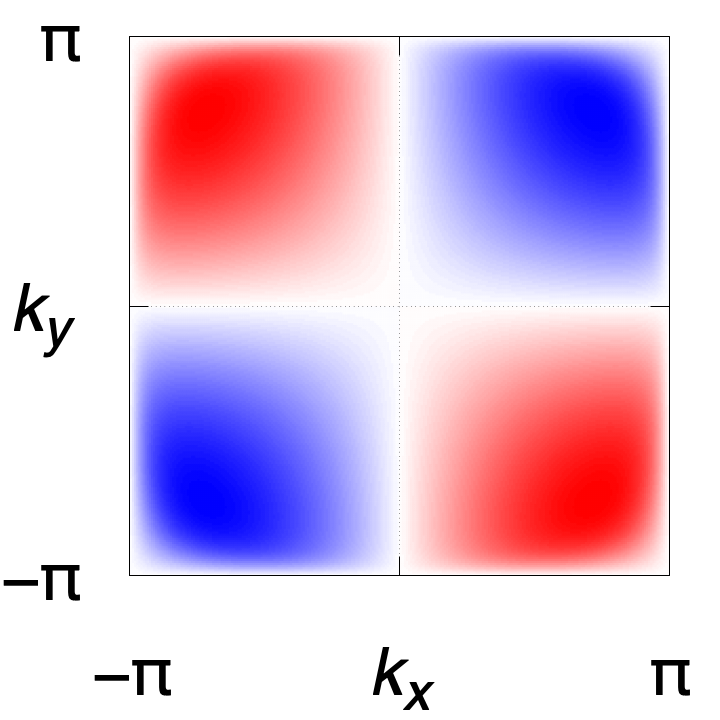}
  & \parbox{40mm}{Phase-modulated (FF) \\ {[Sec.~\ref{sec:GL_square2}, Fig.~\ref{fig:phase_GL_square2}]}} \\ \hline
  \parbox{26mm}{(b) continuum}
  & \includegraphics[valign=c, width=32mm]{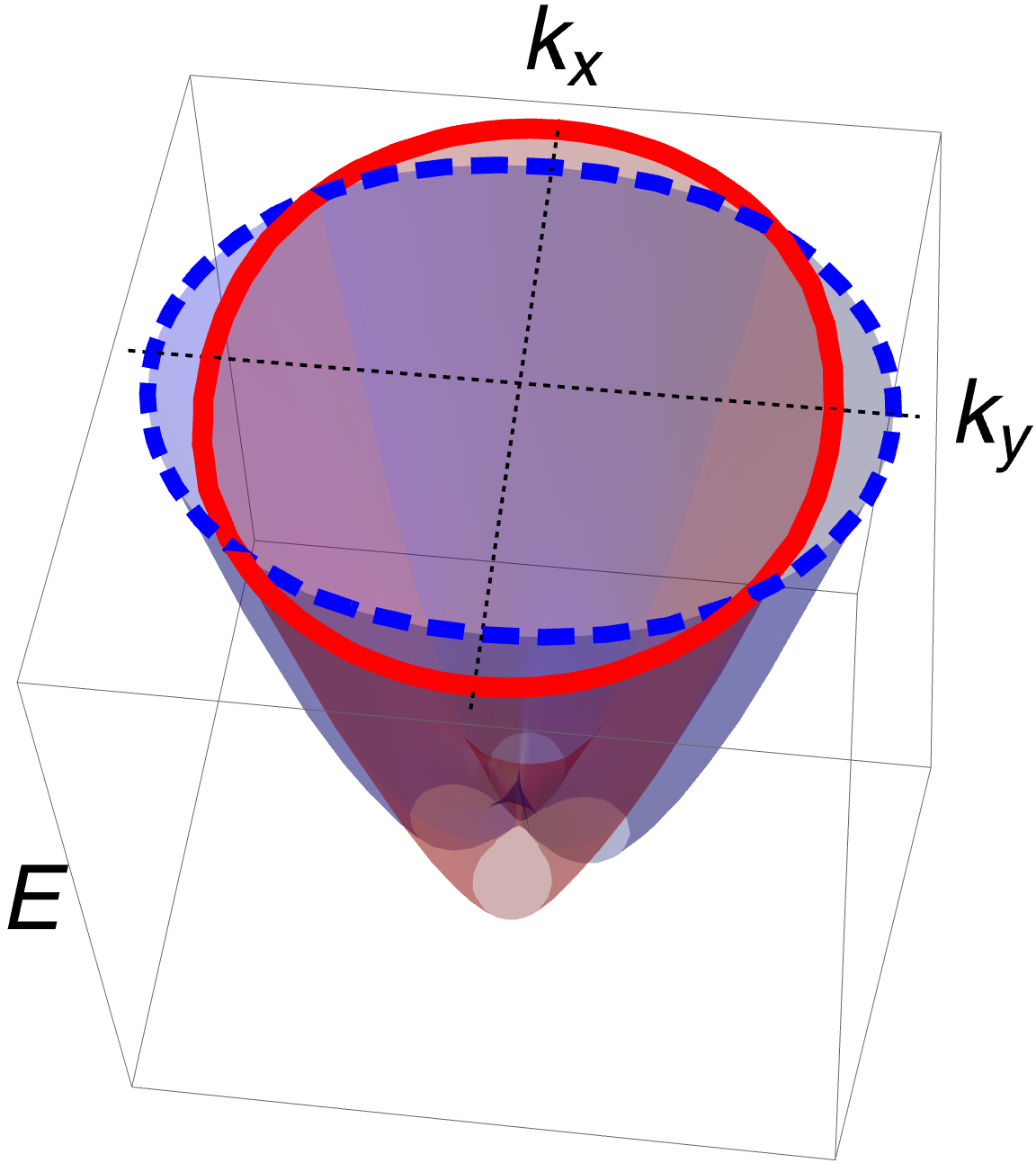}
  & \parbox{40mm}{$- \sqrt{2} h (k_{nx}^2 - k_{ny}^2) \hat{\sigma}_z$ \\ {[Eq.~\eqref{eq:Hamiltonian_continuum_mag}]}}
  & \includegraphics[valign=c, width=28mm]{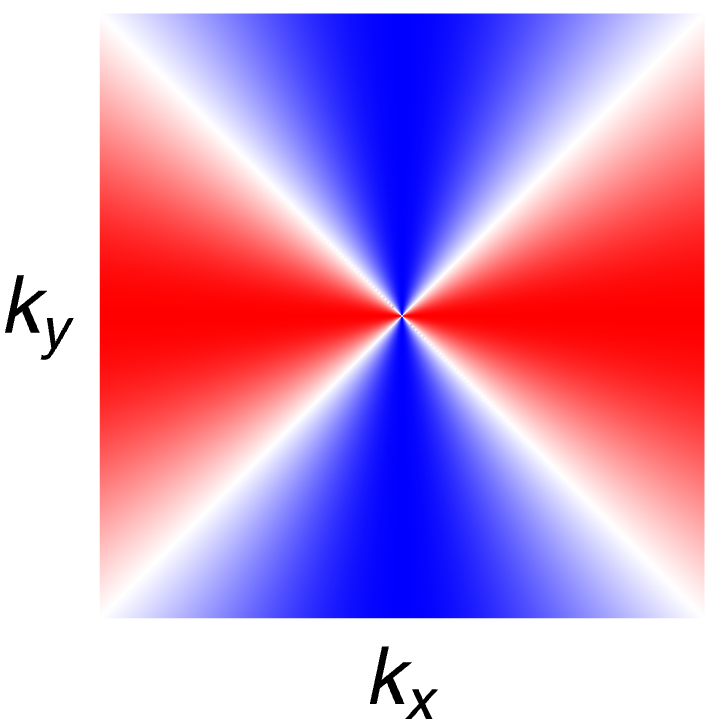}
  & \parbox{40mm}{Amplitude-modulated \\ (LO or bidirectional PDW) \\ {[Sec.~\ref{sec:GL_continuum_mag_d}, Fig.~\ref{fig:phase_GL_continuum_dS}(b)]}} \\ \hline
  \parbox{26mm}{(c1) conventional square lattice ($1\bm{k}$-$d$-wave)}
  & \multirow{2}{*}{\includegraphics[valign=c, width=32mm]{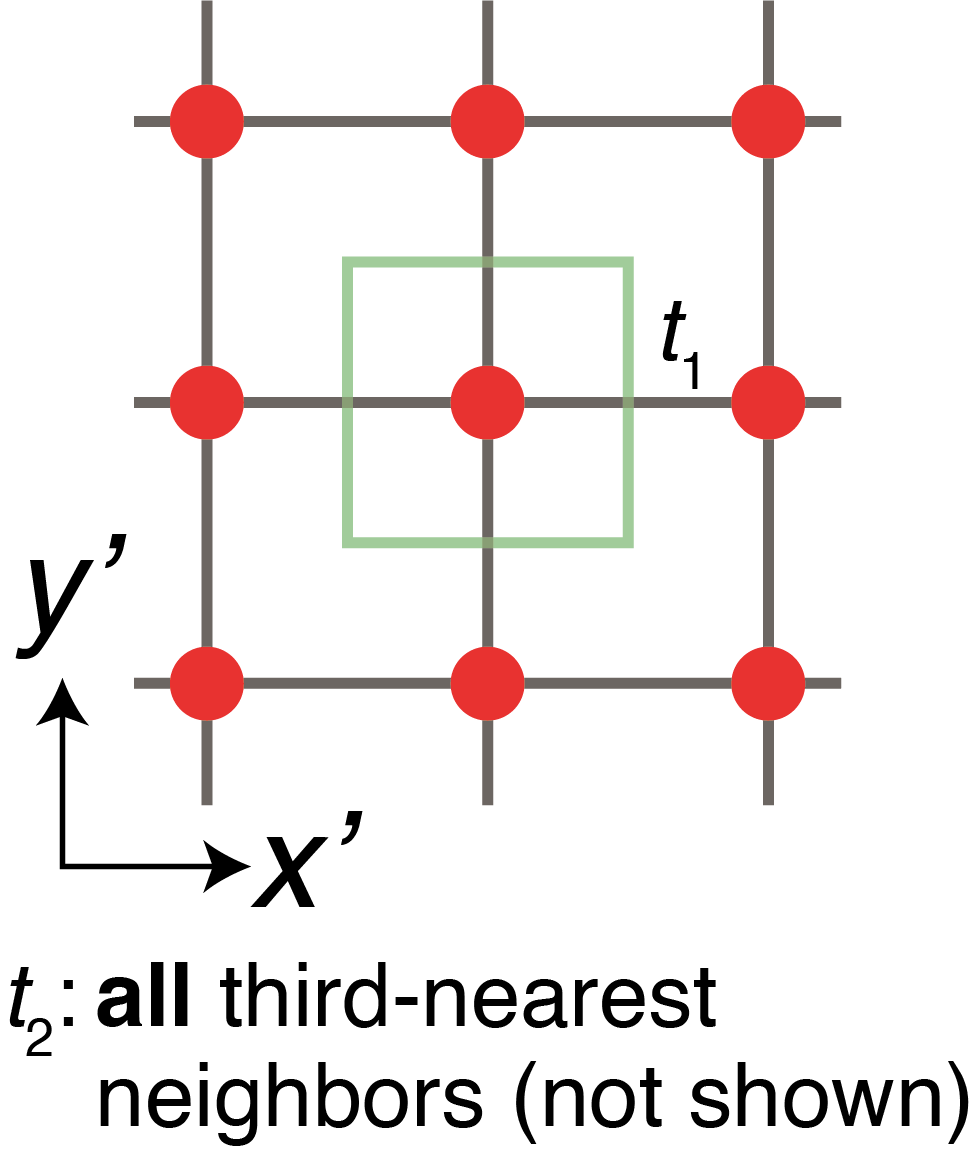}}
  & \parbox{40mm}{$- h (\cos k_x' - \cos k_y') \hat{\sigma}_z$ \\ {[Eq.~\eqref{eq:Hamiltonian_square1_mag_1}]}}
  & \includegraphics[valign=c, width=28mm]{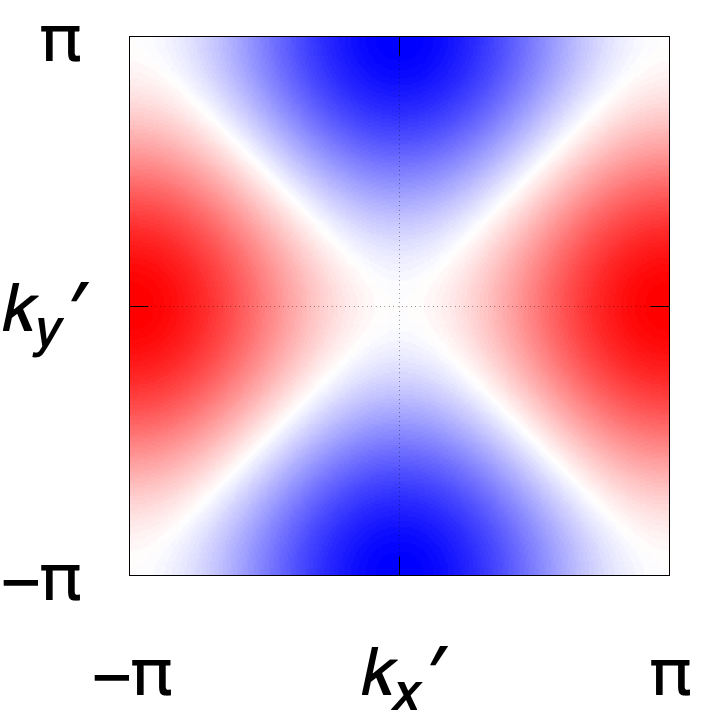}
  & \parbox{40mm}{Amplitude-modulated \\ (LO or bidirectional PDW) \\ {[Sec.~\ref{sec:GL_square1}, Fig.~\ref{fig:phase_GL_square1}(a)]}} \\ \cline{3-5}
  \parbox{26mm}{(c2) conventional square lattice ($2\bm{k}$-$d$-wave)}
  & & \parbox{40mm}{$- h [\cos(2k_x') - \cos(2k_y')] \hat{\sigma}_z$ \\ {[Eq.~\eqref{eq:Hamiltonian_square1_mag_2}]}}
  & \includegraphics[valign=c, width=28mm]{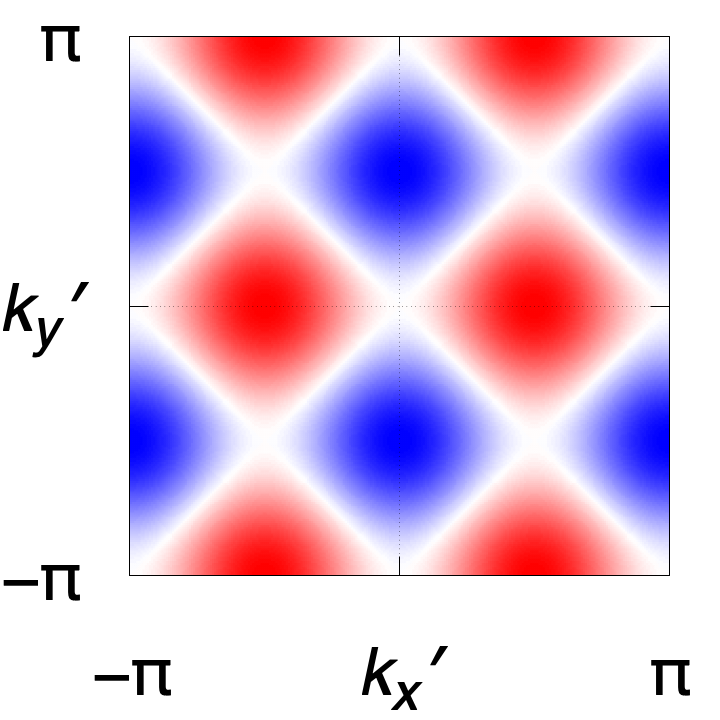}
  & \parbox{40mm}{Phase-modulated (FF) \\ {[Sec.~\ref{sec:GL_square1}, Fig.~\ref{fig:phase_GL_square1}(b)]}} \\ \hline\hline
 \end{tabular}
\end{table*}

\section{Models and electronic structures}
\label{sec:model}
In this section, we introduce the three 2D models discussed in this paper, based on which the stability of superconducting state will be investigated in the next section.
For all the models, the noninteracting Hamiltonian is given by the form,
\begin{equation}
 H_0 = \sum_{\bm{k}} \bm{C}_{\bm{k}}^\dagger \hat{H}_0(\bm{k}) \bm{C}_{\bm{k}},
 \label{eq:Hamiltonian}
\end{equation}
where the momentum-dependent Hamiltonian matrix is composed of two terms,
\begin{equation}
 \hat{H}_0(\bm{k}) = \hat{H}_{\mathrm{kin}}(\bm{k}) + \hat{H}_{\mathrm{mag}}(\bm{k}).
 \label{eq:Hamiltonian_momentum}
\end{equation}
The first term is the kinetic energy, and the second term represents the molecular field of the altermagnetic order or the Zeeman field.
In the following sections, we describe the two-sublattice tetragonal model (Sec.~\ref{sec:model_square2}), the continuum model (Sec.~\ref{sec:model_continuum}), and the conventional square lattice model (Sec.~\ref{sec:model_square1}).

\subsection{Two-sublattice tetragonal model}
\label{sec:model_square2}
\subsubsection{Tight-binding Hamiltonian}
First we introduce the two-sublattice tetragonal model for altermagnets, as shown in Table~\ref{tab:models}(a) (cf. Refs.~\cite{Maier2023, Roig2024}).
This is a simplified model of $\kappa$-type organic compounds, an early identified system with altermagnetic spin splitting~\cite{Naka2019, Naka2020, Seo2021}.
Since this model contains two sublattices in the unit cell, the operator in Eq.~\eqref{eq:Hamiltonian} is a four-component vector for each $\bm{k}$ written as
\begin{equation}
 \bm{C}_{\bm{k}} = [c_{\bm{k}, a \uparrow}, c_{\bm{k}, a \downarrow}, c_{\bm{k}, b \uparrow}, c_{\bm{k}, b \downarrow}]^{\mathrm{T}},
\end{equation}
where $c_{\bm{k}, l s}$ is the annihilation operator of electrons carrying momentum $\bm{k}$ and spin $s$ on the sublattice $l$ ($= a, b$).
The kinetic energy [the first term in Eq.~\eqref{eq:Hamiltonian_momentum}] is composed of the nearest-neighbor intersublattice hopping ($t_1$) and third-nearest-neighbor intrasublattice hopping $(t_2)$ on a square lattice [see the second column in Table~\ref{tab:models}(a)]:
\begin{align}
 \hat{H}_{\mathrm{kin}}(\bm{k}) &=
 - 4t_1 \cos\frac{k_x}{2} \cos\frac{k_y}{2} \hat{\tau}_1 \otimes \hat{\sigma}_0 \notag \\
 &\quad - 2t_2 (\cos k_x \cos k_y \hat{\tau}_0 - \sin k_x \sin k_y \hat{\tau}_z) \otimes \hat{\sigma}_0,
 \label{eq:Hamiltonian_square2_kin}
\end{align}
where the lattice constants are set to unity.
Here $\hat{\tau}_i$ and $\hat{\sigma}_j$ are the Pauli matrices for the sublattice and spin spaces, respectively.
On the other hand, the molecular field of the altermagnetic order [the second term in Eq.~\eqref{eq:Hamiltonian_momentum}] breaking time-reversal symmetry is given by
\begin{equation}
 \hat{H}_{\mathrm{mag}} =
 - h \hat{\tau}_z \otimes \hat{\sigma}_z,
 \label{eq:Hamiltonian_square2_mag}
\end{equation}
which does not depend on the momentum $\bm{k}$, but does on the two sublattices due to $\hat{\tau}_z$.
Since the model does not contain spin--orbit coupling, we choose, without loss of generality, the molecular field along the $z$ axis.

Note that the third-nearest-neighbor hopping $t_2$ in Eq.~\eqref{eq:Hamiltonian_square2_kin} has an anisotropic structure between the two sublattices.
Reflecting the different orientations of the sublattices as in the $\kappa$-type organic compounds where the role of anisotropic hoppings was elucidated~\cite{Naka2019, Naka2020, Seo2021}, the hoppings are introduced only along the $[11]$ ($[1\overline{1}]$) direction for the $a$ ($b$) sublattice; see the second column in Table~\ref{tab:models}(a).
This anisotropy causes the spin splitting in the energy band structure, when the molecular field [Eq.~\eqref{eq:Hamiltonian_square2_mag}] is nonzero.
We use the hopping parameters $(t_1, \, t_2) = (1, \, 0.2)$ throughout this paper.
For simplicity, we neglect next-nearest-neighbor hoppings and treat this minimal model~\cite{Naka2019, Roig2024} to describe the altermagnetic characters.
At the end of Sec.~\ref{sec:GL}, we will briefly address how incorporating these hoppings influences superconductivity.

The Hamiltonian, in the form above, does not satisfy the periodicity with respect to the translation $\bm{k} \to \bm{k} + \bm{K}$ with $\bm{K}$ being a reciprocal lattice vector.
We thus perform a unitary transformation by the following unitary matrix
\begin{equation}
 \hat{U}_{\mathrm{site}}(\bm{k}) =
 \begin{bmatrix}
  \ee^{\ii k_y / 2} & 0 \\
  0 & \ee^{\ii k_x / 2}
 \end{bmatrix}_\tau
 \otimes \hat{\sigma}_0,
\end{equation}
to restore the periodicity of the Brillouin zone (BZ).

\subsubsection{Symmetry and energy band structure}
When the altermagnetic term [Eq.~\eqref{eq:Hamiltonian_square2_mag}] is absent, the model belongs to a 2D magnetic point group $4mm1'$ that is generated by fourfold rotation ($C_{4z}$), mirror (glide) with respect to the $x$ axis ($M_x$), and time-reversal ($T$) symmetries~%
\footnote{%
We here do not take into account symmetry operators that acts only on spin--orbit-coupling-free systems.
In other words, we do not consider a spin point group.
}:
\begin{subequations}
 \label{eq:symmetry_relation}
 \begin{align}
  \hat{U}_{C_{4z}} \hat{H}_0(\bm{k}) \hat{U}_{C_{4z}}^\dagger &= \hat{H}_0(-k_y, k_x), \\
  \hat{U}_{M_x} \hat{H}_0(\bm{k}) \hat{U}_{M_x}^\dagger &= \hat{H}_0(-k_x, k_y), \\
  \hat{U}_T \hat{H}_0(\bm{k})^* \hat{U}_T^\dagger &= \hat{H}_0(-\bm{k}).
 \end{align}
\end{subequations}
Here the unitary matrices representing the symmetry operations are defined by
\begin{subequations}
 \label{eq:symmetry_unitary}
 \begin{align}
  \hat{U}_{C_{4z}} &:= \hat{\tau}_x \otimes \exp\left(-\ii \frac{\pi}{4} \hat{\sigma}_z\right), \displaybreak[2] \\
  \hat{U}_{M_x} &:= \hat{\tau}_x \otimes (-\ii\hat{\sigma}_x), \displaybreak[2] \\
  \hat{U}_T &:= \hat{\tau}_0 \otimes \ii\hat{\sigma}_y.
 \end{align}
\end{subequations}
When the altermagnetic order is turned on, some of the symmetries (including time-reversal symmetry $T$) are broken, and the remaining symmetry operations are represented by $4'mm'$, which is a subgroup of $4mm1'$.

\begin{figure}
 \includegraphics[width=.75\linewidth, pagebox=artbox]{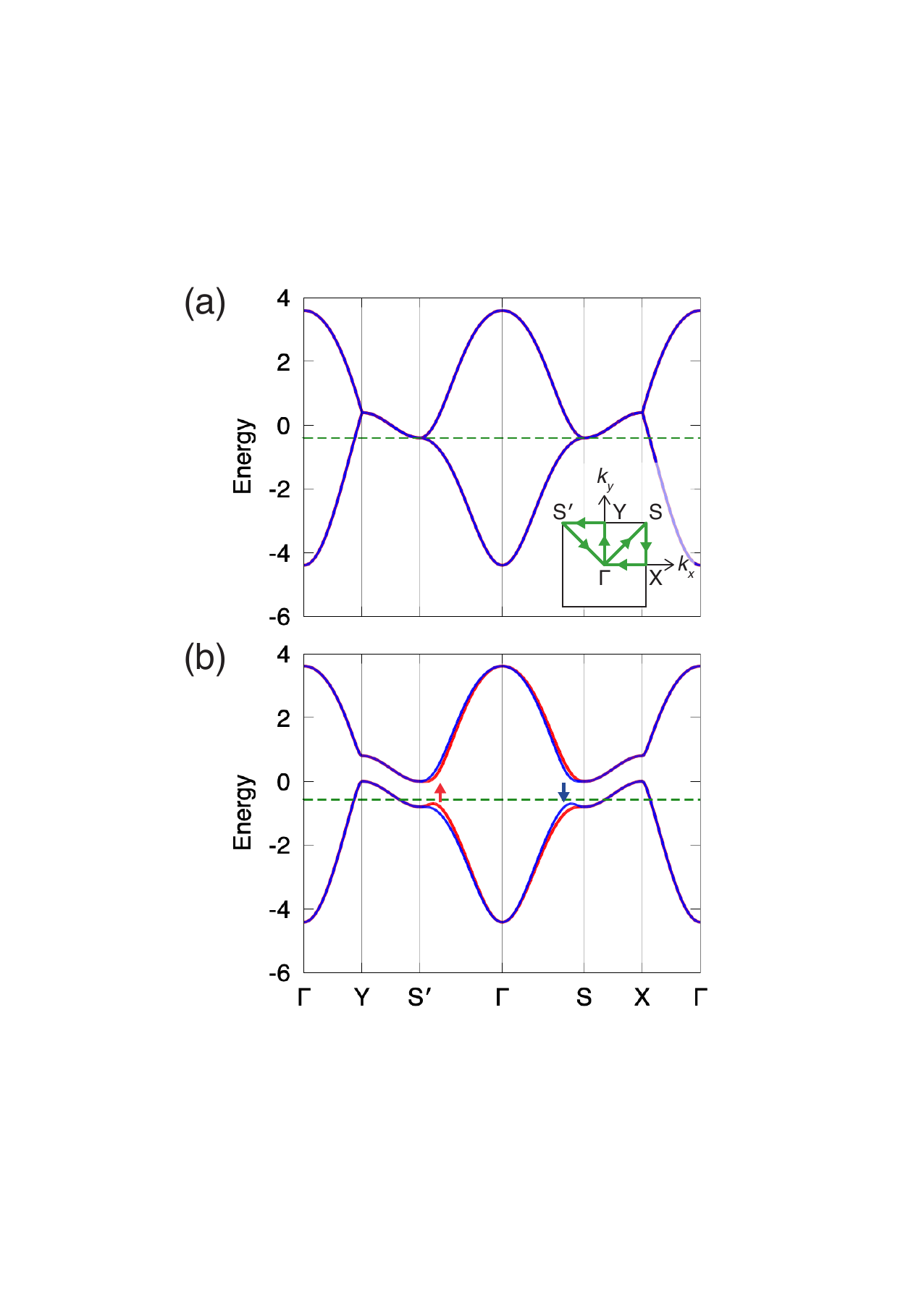}
 \caption{Energy band structures for (a) $h = 0$ and (b) $0.4$. Energy bands of up- and down-spin electrons are represented by the red and blue lines, respectively. The green dashed lines represent the Fermi level for the electron density $n = 0.85$ per site. The inset in (a) shows the high-symmetry path in the 2D BZ. We use symbols ($\Gamma$, X, Y, S) of the primitive orthorhombic lattice since the fourfold rotation symmetry is broken in the altermagnetic state.}
 \label{fig:energy_band}
\end{figure}

By diagonalizing the noninteracting Hamiltonian $\hat{H}_0(\bm{k})$, we obtain the energy band dispersion.
Figures~\ref{fig:energy_band}(a) and \ref{fig:energy_band}(b) show the energy band structures for $h = 0$ and $h = 0.4$, respectively.
When the molecular field $h$ is finite, the spin splitting appears at general $\bm{k}$ points, except along the $k_{x, y}$ axes and the BZ boundary [see also the fourth column in Table~\ref{tab:models}(a)].
In the following discussions, we focus on a hole-doped case with the electron density $n = n_\uparrow + n_\downarrow$ per site being $0.85$ since our previous study has indicated that the FFLO state is more likely in the hole-doped regime~\cite{Sumita2023}.
The green dashed lines in Figs.~\ref{fig:energy_band}(a) and \ref{fig:energy_band}(b) show the Fermi level for $n = 0.85$.

\subsection{Continuum model}
\label{sec:model_continuum}
Next, we introduce the isotropic continuum model with the Zeeman field, as shown in Table~\ref{tab:models}(b).
Since this model does not contain the sublattice degree of freedom, the annihilation operator in Eq.~\eqref{eq:Hamiltonian} is now given by $\bm{C}_{\bm{k}} = [c_{\bm{k}, \uparrow}, c_{\bm{k}, \downarrow}]^{\mathrm{T}}$, and the Hamiltonian is written as
\begin{align}
 \hat{H}_{\mathrm{kin}}(\bm{k}) &= \frac{\bm{k}^2}{2m} \hat{\sigma}_0,
 \label{eq:Hamiltonian_continuum_kin} \\
 \hat{H}_{\mathrm{mag}}(\bm{k}) &= - h(\bm{k}_n) \hat{\sigma}_z,
 \label{eq:Hamiltonian_continuum_mag}
\end{align}
where the kinetic term is a parabolic dispersion.
The Zeeman field in Eq.~\eqref{eq:Hamiltonian_continuum_mag} depends on the direction of the momentum $\bm{k}_n = \bm{k} / |\bm{k}|$.
In Sec.~\ref{sec:GL_continuum}, we discuss the GL theory for two types of the momentum dependence, namely, a uniform Zeeman field
\begin{equation}
 h(\bm{k}_n) = h,
 \label{eq:Hamiltonian_continuum_mag_s}
\end{equation}
and a $d$-wave Zeeman field (cf. Refs.~\cite{Zhang2024, Mukasa2025, Sim2024, Hong2025, Iorsh2025, Hu2025_arXiv_1, Hu2025_arXiv_2})
\begin{equation}
 h(\bm{k}_n) = \sqrt{2} h (k_{nx}^2 - k_{ny}^2),
 \label{eq:Hamiltonian_continuum_mag_d}
\end{equation}
with $h$ being a magnitude of the magnetic order or magnetic field.
Particularly in the $d$-wave Zeeman field~\eqref{eq:Hamiltonian_continuum_mag_d}, the energy dispersion exhibits anisotropic spin splitting resembling that in altermagnets, as shown in Table~\ref{tab:models}(b).

\begin{figure}
 \includegraphics[width=.9\linewidth, pagebox=artbox]{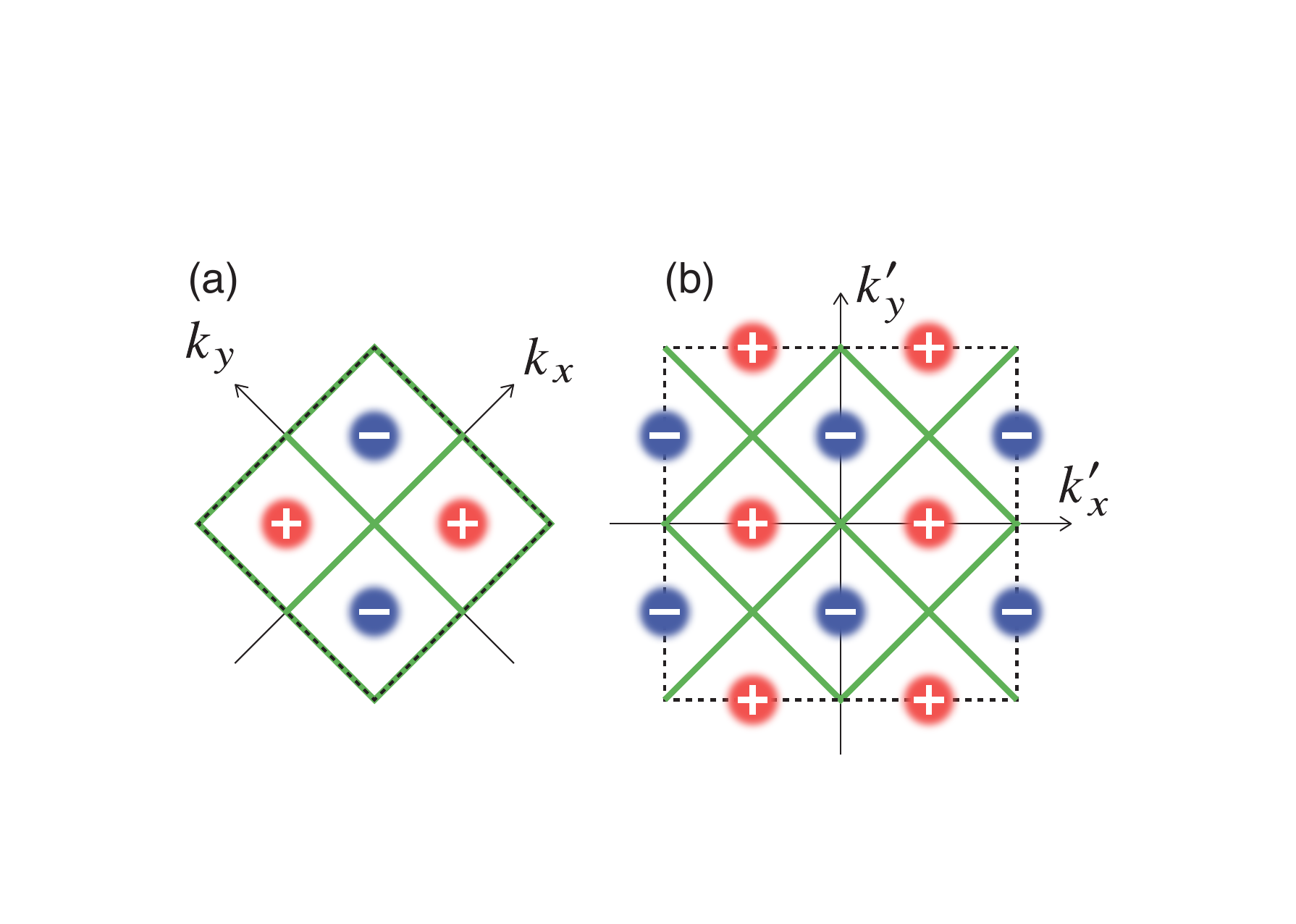}
 \caption{The first BZs associated with the spin splitting structures in (a) the two-sublattice tetragonal model and (b) the conventional square lattice model with $2\bm{k}$-$d$-wave Zeeman field. In the altermagnetic state, energy dispersions have $d$-wave type spin splitting, which is illustrated by the plus and minus signs. Green lines represent the nodes, where the up- and down-spin bands are degenerate even in the presence of the magnetic order, and the dashed lines show the first BZ boundaries.}
 \label{fig:BZ_nodes}
\end{figure}

\subsection{Conventional square lattice model}
\label{sec:model_square1}
Finally, we introduce the conventional square lattice model with the anisotropic Zeeman term [Table~\ref{tab:models}(c)].
In the two-sublattice tetragonal model (Sec.~\ref{sec:model_square2}), the $d$-wave type spin splitting emerges when the anisotropic third-nearest-neighbor hopping $t_2$ and the molecular field $h$ alternating on the two sublattices are finite.
Here, to mimic the spin splitting, we consider a momentum-dependent Zeeman field.
The annihilation operator in Eq.~\eqref{eq:Hamiltonian} is given by $\bm{C}_{\bm{k}'} = [c_{\bm{k}', \uparrow}, c_{\bm{k}', \downarrow}]^{\mathrm{T}}$, and the Hamiltonian is
\begin{align}
 \hat{H}_{\mathrm{kin}}(\bm{k}') &=
 -2t_1 (\cos k_x' + \cos k_y') \hat{\sigma}_0 \notag \\
 &\quad - 2t_2 (\cos 2k_x' + \cos 2k_y') \hat{\sigma}_0,
 \label{eq:Hamiltonian_square1_kin} \\
 \hat{H}_{\mathrm{mag}}(\bm{k}') &=
 - h(\bm{k}') \hat{\sigma}_z,
 \label{eq:Hamiltonian_square1_mag}
\end{align}
which only have the spin degree of freedom described by the spin space Pauli matrices $\hat{\sigma}_j$.
We use the primed momentum $\bm{k}'$ to distinguish the momentum $\bm{k}$ in the two-sublattice tetragonal model [Eqs.~\eqref{eq:Hamiltonian_square2_kin} and \eqref{eq:Hamiltonian_square2_mag}], where the sublattice degree of freedom was described by the Pauli matrices $\hat{\tau}_i$.
Considering the folding of the BZ, $(k_x', k_y')$ is identified as $\frac{1}{2} (k_x - k_y, k_x + k_y)$ (see Fig.~\ref{fig:BZ_nodes}).
The kinetic term \eqref{eq:Hamiltonian_square1_kin} contains the transfer integral $t_2$ for all third-nearest neighbors, in contrast to the anisotropic $t_2$ in the two-sublattice tetragonal model [Eq.~\eqref{eq:Hamiltonian_square2_kin}].

Equation~\eqref{eq:Hamiltonian_square1_mag}, which has the same form with Eq.~\eqref{eq:Hamiltonian_continuum_mag} in the continuum model, represents the momentum-dependent Zeeman field.
We consider two forms of $h(\bm{k}')$; the first one is given by
\begin{equation}
 h(\bm{k}') = h (\cos k_x' - \cos k_y'),
 \label{eq:Hamiltonian_square1_mag_1}
\end{equation}
which gives the $d$-wave type structure of spin splitting as shown in the fourth column in Table~\ref{tab:models}(c1).
While this form of $h(\bm{k}')$ was discussed in previous studies~\cite{Chakraborty2024, Chakraborty2024_arXiv_1, Chakraborty2024_arXiv_2}, strictly speaking, it generates a spin-split structure different from that in the two-sublattice tetragonal model, in which the nodes (green lines) are located at the zone boundaries $k_i = \pm \pi$ ($i = x, y$) as well as the interior $k_i = 0$ [see the fourth column in Table~\ref{tab:models}(a)].
Therefore here we introduce another form of $h(\bm{k}')$ with momentum doubling,
\begin{equation}
 h(\bm{k}') = h [\cos(2k_x') - \cos(2k_y')].
 \label{eq:Hamiltonian_square1_mag_2}
\end{equation}
The splitting structure in the extended BZ of the conventional square lattice model with one sublattice is illustrated in the fourth column in Table~\ref{tab:models}(c2), which reproduces the structure in the two-sublattice tetragonal model; the relationship between spin splitting structures are summarized in Fig.~\ref{fig:BZ_nodes}.
To distinguish the two cases, we will refer to Eqs.~\eqref{eq:Hamiltonian_square1_mag_1} and \eqref{eq:Hamiltonian_square1_mag_2} as the $1\bm{k}$- and $2\bm{k}$-$d$-wave Zeeman fields, respectively.

\section{Ginzburg--Landau theory}
\label{sec:GL}
In this section, to discuss the spatial structure of finite-momentum superconductivity in altermagnets, we construct the GL theory based on the microscopic Hamiltonian with superconducting instability.
First, we show the general formalism of the GL theory in Sec.~\ref{sec:GL_formalism}.
Next, in Secs.~\ref{sec:GL_square2}--\ref{sec:GL_square1}, we apply the theory to the three models introduced in Sec.~\ref{sec:model}.
The source codes used for the numerical calculations are available in Ref.~\cite{SourceCode}.

\subsection{Formalism}
\label{sec:GL_formalism}
We here introduce the microscopic derivation of the GL theory~\cite{Gorkov1964, Ren1995, Feder1997}.
Let us begin with the Hamiltonian using the superconducting MF approximation with an order parameter in the matrix form $\hat{\Delta}(\bm{x}, \bm{x}')$,
\begin{align}
 H_{\mathrm{MF}} &= H_0 + \frac{1}{2} \sum_{\bm{x}, \bm{x}'} \sum_{\zeta, \zeta'} \left[ \Delta(\bm{x}, \bm{x}')_{\zeta, \zeta'} c_{\bm{x}, \zeta}^\dagger c_{\bm{x}', \zeta'}^\dagger + \text{H.c.} \right],
\end{align}
where $\bm{x}$ and $\bm{x}'$ represent real-space coordinates, and we label the internal degrees of freedom with $\zeta$.
We then perform Fourier transformation with respect to the relative coordinate $\bm{r} = \bm{x} - \bm{x}'$:
\begin{subequations}
 \begin{align}
  \hat{\Delta}(\bm{x}, \bm{x}') &= \frac{1}{N} \sum_{\bm{k}} \ee^{\ii \bm{k} \cdot \bm{r}} \hat{\Delta}(\bm{R}, \bm{k}), \\
  \hat{\Delta}(\bm{R}, \bm{k}) &= \sum_{\bm{r}} \ee^{-\ii \bm{k} \cdot \bm{r}} \hat{\Delta}\left(\bm{R} + \frac{\bm{r}}{2}, \bm{R} - \frac{\bm{r}}{2}\right),
  \label{eq:order_parameter_Wigner}
 \end{align}
\end{subequations}
where $\bm{R} = (\bm{x} + \bm{x}') / 2$ is the COM coordinate and $N$ is the number of unit cells.
Equation~\eqref{eq:order_parameter_Wigner} is called the Wigner representation of the order parameter matrix~\cite{Wigner1932}.
We here consider a separable form $\hat{\Delta}(\bm{R}, \bm{k}) = \eta(\bm{R}) \hat{\phi}(\bm{k})$ of the order parameter, where the magnitude $\eta(\bm{R})$ satisfies the MF equation
\begin{equation}
 \eta(\bm{R}) = - \frac{TV}{N} \sum_{k} \tr\left[ \hat{\phi}(\bm{k})^\dagger \hat{F}(\bm{R}, \bm{k}; \omega_n) \right],
 \label{eq:mean-field}
\end{equation}
with an effective interaction $V > 0$.
We use the fermionic Matsubara frequency $\omega_n = (2n + 1)\pi T$ at temperature $T$ and the abbreviation $\sum_{k} = \sum_{\omega_n} \sum_{\bm{k}}$.
$\hat{\phi}(\bm{k})$ provides the matrix basis of the order parameter satisfying $\hat{\phi}(\bm{k}) = - \hat{\phi}(-\bm{k})^{\mathrm{T}}$, where the overall factor is chosen so that the normalization condition
\begin{equation}
 \frac{1}{N} \sum_{\bm{k}} \tr\left[\hat{\phi}(\bm{k})^\dagger \hat{\phi}(\bm{k})\right]
 = (\text{spin degree of freedom}) = 2
\end{equation}
is satisfied.
The Wigner representation $F(\bm{R}, \bm{k}; \omega_n)_{\zeta, \zeta'}$ is obtained by Fourier transformation of the anomalous Green's function $F(\bm{x}, \bm{x}'; \tau)_{\zeta, \zeta'} = - \braket{T_\tau c_{\bm{x}, \zeta}(\tau) c_{\bm{x}', \zeta'}}$ with respect to the relative coordinate $\bm{r}$ and the imaginary time $\tau$:
\begin{equation}
 \hat{F}(\bm{R}, \bm{k}; \omega_n)
 = \sum_{\bm{r}} \ee^{-\ii\bm{k} \cdot \bm{r}} \int_{0}^{\beta} \mathrm{d}\tau \, \ee^{\ii\omega_n \tau} \hat{F}\left(\bm{R} + \frac{\bm{r}}{2}, \bm{R} - \frac{\bm{r}}{2}; \tau\right).
\end{equation}

Next, we derive the GL equation and free energy.
In the following, we assume that the magnitude of the order parameter $\eta(\bm{R})$ is sufficiently small and slowly varies with respect to $\bm{R}$.
We expand the MF Eq.~\eqref{eq:mean-field} in terms of $\eta$ and its spatial derivative with the use of the Gor'kov equations,
\begin{subequations}
 \label{eq:Gorkov}
 \begin{align}
  \hat{G}(\bm{x}, \bm{x}'; \omega_n) &= \hat{G}_0(\bm{x}, \bm{x}'; \omega_n) \notag \\
  &\quad - \sum_{\bm{x}_1, \bm{x}_2} \hat{G}_0(\bm{x}, \bm{x}_1; \omega_n) \hat{\Delta}(\bm{x}_1, \bm{x}_2) \hat{F}(\bm{x}_2, \bm{x}'; \omega_n)^*, \\
  \hat{F}(\bm{x}, \bm{x}'; \omega_n) &= - \sum_{\bm{x}_1, \bm{x}_2} \hat{G}_0(\bm{x}, \bm{x}_1; \omega_n) \hat{\Delta}(\bm{x}_1, \bm{x}_2) \hat{G}(\bm{x}_2, \bm{x}'; \omega_n)^*,
 \end{align}
\end{subequations}
where $G(\bm{x}, \bm{x}'; \omega_n)_{\zeta, \zeta'}$ is the Fourier component of the normal Green's function $G(\bm{x}, \bm{x}'; \tau)_{\zeta, \zeta'} = - \braket{T_\tau c_{\bm{x}, \zeta}(\tau) c_{\bm{x}', \zeta'}^\dagger}$, and $\hat{G}_0$ is the normal Green's function of free electrons.
After some calculations, we obtain the GL equation,
\begin{align}
 & \alpha \eta(\bm{R}) + 2\beta |\eta(\bm{R})|^2 \eta(\bm{R}) - \kappa \nabla_{\bm{R}}^2 \eta(\bm{R}) \notag \\
 & + \delta \nabla_{\bm{R}}^4 \eta(\bm{R}) + \delta' \partial_X^2 \partial_Y^2 \eta(\bm{R}) \notag \\
 & + (\text{higher-order terms}) = 0,
 \label{eq:GL}
\end{align}
in tetragonal systems [$\bm{R} = (X, Y)$].
For the detailed derivation, see Appendix~\ref{app:GL_microscopic_derivation}.
This equation can be seen as a saddle point solution $\delta \mathcal{F} / \delta \eta^* = 0$ of the free energy functional $\mathcal{F}[\eta(\bm{R})] = \frac{1}{\mathcal{V}} \int \mathrm{d}\bm{R} \, f(\bm{R})$ with $\mathcal{V}$ being the volume of the system.
The GL free energy density per unit cell is then defined by
\begin{equation}
 f = \alpha |\eta|^2 + \beta |\eta|^4 + \kappa |\nabla\eta|^2
 + \delta |\nabla^2\eta|^2 + \delta' |\partial_X \partial_Y \eta|^2,
 \label{eq:GL_free_energy_density}
\end{equation}
up to the fourth order of $\eta$ and spatial derivatives $\partial_i$ (we omit the $\bm{R}$ dependence for simplicity).
The GL coefficients of the uniform terms are calculated as
\begin{align}
 \alpha
 &= \frac{1}{V} - \frac{T}{N} \sum_{k} \tr\left[\hat{X}(\bm{k}, \bm{0}; \omega_n)\right],
 \label{eq:GL_coefficient_alpha} \displaybreak[2] \\
 \beta
 &= \frac{T}{2N} \sum_{k} \tr\left[\hat{X}(\bm{k}, \bm{0}; \omega_n)^2\right],
 \label{eq:GL_coefficient_beta}
\end{align}
where we define
\begin{equation}
 \hat{X}(\bm{k}, \bm{Q}; \omega_n)
 := \hat{\phi}(\bm{k})^\dagger \hat{G}_0\left(\bm{k}_+; \omega_n\right)
 \hat{\phi}(\bm{k}) \hat{G}_0\left(\bm{k}_-; \omega_n\right)^*,
 \label{eq:X_k_Q}
\end{equation}
with $\bm{k}_\pm = \pm \bm{k} + \bm{Q}$ and the noninteracting Green's function $\hat{G}_0(\bm{k}; \omega_n) = [ (\ii\omega_n + \mu) \hat{\bm{1}} - \hat{H}_0(\bm{k}) ]^{-1}$ for the chemical potential $\mu$.
The coefficients of the gradient terms are given by
\begin{align}
 \kappa
 &= - c^{(2)}_{XX} =  - c^{(2)}_{YY},
 \label{eq:GL_coefficient_kappa} \\
 \delta
 &= c^{(2)}_{XXXX} = c^{(2)}_{YYYY},
 \label{eq:GL_coefficient_delta} \\
 \delta'
 &= 6c^{(2)}_{XXYY} - 2c^{(2)}_{XXXX},
 \label{eq:GL_coefficient_delta'}
\end{align}
where
\begin{align}
 c^{(2)}_{\mu_1 \dots \mu_n}
 &:= - \frac{1}{n! (2\ii)^n} \frac{T}{N} \notag \\
 &\quad \times \sum_{k} \frac{\partial^n}{\partial Q_{\mu_1} \dotsb \partial Q_{\mu_n}}
 \tr\left[ \hat{X}(\bm{k}, \bm{Q}; \omega_n) \right] \bigg|_{\bm{Q} = \bm{0}}.
 \label{eq:c2_diff}
\end{align}
Note that $\delta'$ corresponds to the anisotropic term allowed in the tetragonal system.
In the formulation of Eqs.~\eqref{eq:GL_coefficient_alpha}, \eqref{eq:GL_coefficient_beta}, and \eqref{eq:GL_coefficient_kappa}--\eqref{eq:GL_coefficient_delta'}, we do not include the effect of vector potential since we consider cases without an external magnetic field.

\begin{figure}
 \includegraphics[width=.8\linewidth, pagebox=artbox]{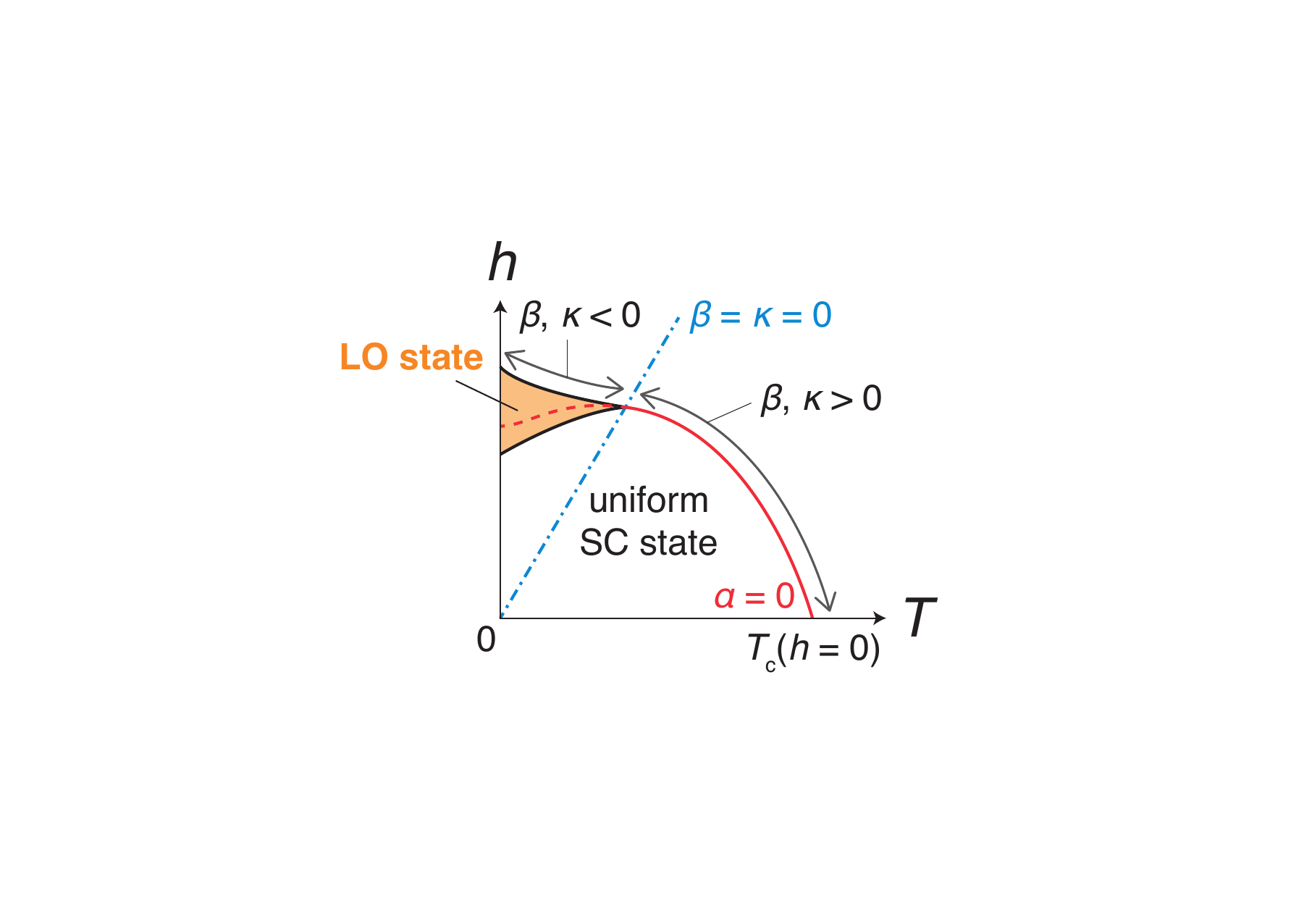}
 \caption{Schematic phase diagram in a uniform Zeeman field $h$. The zeros of the GL coefficients $\alpha$ and $\beta$ ($\kappa$) are depicted by red and blue lines, respectively. The amplitude-modulated LO state (orange shaded area) is stabilized in the high field region, where both $\beta$ and $\kappa$ are negative.}
 \label{fig:phase_GL_uniform_Zeeman}
\end{figure}

Before showing the results analyzing the derived GL coefficients in the following subsections, let us briefly review how the GL theory provides the interpretation of the FFLO state in a uniform Zeeman field~\cite{Buzdin1997, Agterberg2001}, on the basis of GL free energy in Eq.~\eqref{eq:GL_free_energy_density}.
Here let $h$ be the magnitude of the uniform Zeeman field.
When $h$ is small, the superconducting transition temperature $T_{\mathrm{c}}(h)$ is determined by the condition $\alpha = 0$ (see Fig.~\ref{fig:phase_GL_uniform_Zeeman}).
Around the transition temperature, both $\beta$ and $\kappa$ are positive.
Then when we trace along the transition to higher magnetic field region, $\beta$ and $\kappa$ change their sign to negative almost \textit{simultaneously} (we can show that the sign changes exactly coincide in the isotropic continuum model; see Sec.~\ref{sec:GL_continuum}).
Because of the negative $\kappa$, a spatially modulated superconducting state with finite COM momentum is more stable than a uniform state.
Assuming a \textit{phase-modulated} FF state $\eta(\bm{R}) = \eta_0 \ee^{2\ii\bm{Q} \cdot \bm{R}}$, where $\eta_0$ is the solution of the self-consistent gap equation, the COM momentum $\bm{Q}$ that minimizes the free energy is given by
\begin{equation}
 2\bm{Q} =
 \begin{cases}
  \left( \pm \sqrt{\frac{-\kappa}{2\delta}}, \, 0 \right), \, \left( 0, \, \pm \sqrt{\frac{-\kappa}{2\delta}} \right)
  & \delta' \geq 0, \\
  \left( \pm \sqrt{\frac{-\kappa}{4\delta + \delta'}}, \, \pm \sqrt{\frac{-\kappa}{4\delta + \delta'}} \right)
  & \delta' < 0,
 \end{cases}
 \label{eq:COM_momentum}
\end{equation}
where we assume $\delta > 0$ and $4\delta + \delta' > 0$ (indeed, these inequalities are satisfied in our numerical calculations around the FFLO region presented in Sec.~\ref{sec:GL_square2}).
At the optimal COM momentum, we obtain the free energy
\begin{equation}
 \mathcal{F}_{\mathrm{FF}} = \tilde{\alpha} |\eta_0|^2 + \beta |\eta_0|^4,
 \label{eq:free_energy_FF}
\end{equation}
where we define the renormalized coefficient
\begin{equation}
 \tilde{\alpha} :=
 \begin{cases}
  \alpha - \frac{\kappa^2}{4\delta}
  & \delta' \geq 0, \\
  \alpha - \frac{\kappa^2}{4\delta + \delta'}
  & \delta' < 0.
 \end{cases}
 \label{eq:GL_coefficient_alpha_tilde}
\end{equation}
In an \textit{amplitude-modulated} LO state $\eta(\bm{R}) = \sqrt{2} \eta_0 \cos(2\bm{Q} \cdot \bm{R})$, on the other hand, the free energy is calculated as
\begin{equation}
 \mathcal{F}_{\mathrm{LO}} = \tilde{\alpha} |\eta_0|^2 + \frac{3}{2} \beta |\eta_0|^4,
 \label{eq:free_energy_LO}
\end{equation}
which is different from Eq.~\eqref{eq:free_energy_FF} in the fourth-order terms.
Since $\beta$ is negative when $\kappa$ becomes negative as mentioned above, the LO state is more stabilized; this is the scenario for the case of the uniform Zeeman field.
The key point here is the simultaneous sign change in $\beta$ and $\kappa$.
This situation is not always satisfied; for example, in Refs.~\cite{Agterberg2001, Houzet2006}, it is proposed that the sign changes in $\beta$ and $\kappa$ deviate when the impurity effect is taken into account.

Note that the evaluation here is valid when $\kappa$ is negative but close to zero; for large $|\kappa|$, namely a large COM momentum, the $\mathcal{O}(|\eta_0|^4)$ terms in Eqs.~\eqref{eq:free_energy_FF} and \eqref{eq:free_energy_LO} are corrected by terms with the fourth order in $\eta$ and the second order in the derivatives~\cite{Agterberg2001}.
In addition, when $\beta$ is negative, we need to take into account the sixth-order term in $\eta$, which is not included in Eq.~\eqref{eq:GL_free_energy_density}, to discuss the transition temperature of the LO state.
However, in this study we only consider the terms appearing in Eq.~\eqref{eq:GL_free_energy_density}, which are the minimal set to identify whether a phase-modulated or an amplitude-modulated state is stabilized for a small negative $\kappa$.

\subsection{Two-sublattice tetragonal model}
\label{sec:GL_square2}
In this section, we apply the GL theory to the two-sublattice tetragonal model, and also perform the MF calculations for finding the value of $\bm{Q}$.
We show that the spatially nonuniform superconductivity coexisting with an altermagnetic order prefers the FF state, rather than the LO state that is realized in a uniform Zeeman field.
Later, in the next sections~\ref{sec:GL_continuum} and \ref{sec:GL_square1}, to look for the key ingredients giving rise to such a difference, comparisons will be made with other models with a momentum-dependent Zeeman field that has been used in most of previous studies working on FFLO superconductivity in an altermagnetic state~\cite{Zhang2024, Chakraborty2024, Chakraborty2024_arXiv_2, Sim2024, Hong2025, Mukasa2025, Iorsh2025, Hu2025_arXiv_1, Hu2025_arXiv_2} or a spin-triplet nematic state~\cite{Soto-Garrido2014, Gukelberger2014}.

\begin{figure}
 \includegraphics[width=.9\linewidth, pagebox=artbox]{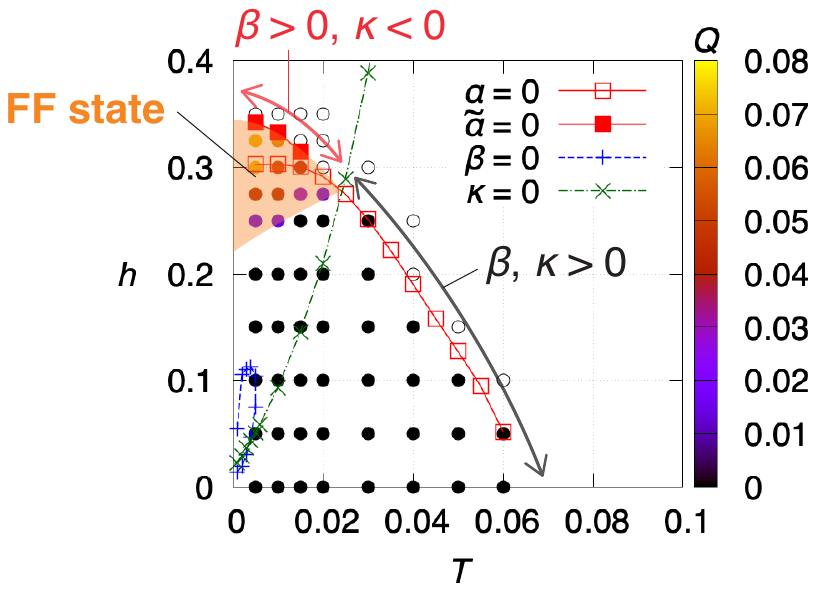}
 \caption{Temperature ($T$) vs altermagnetic molecular field ($h$) phase diagram for the two-sublattice tetragonal model. The zeros of the GL coefficients $\alpha$ ($\tilde{\alpha}$), $\beta$, and $\kappa$ are depicted by red solid, blue dashed, and green dash-dotted lines, respectively. The coefficient $\beta$ is positive in the whole parameters range except in the narrow region enclosed by the blue dashed line. The filled (open) circles represent the superconducting (normal) state obtained by the MF theory, where the color of the filled circles indicates the magnitude of the optimal COM momentum $Q = |\bm{Q}|$. The effective interaction is set to $V = 0.35$.}
 \label{fig:phase_GL_square2}
\end{figure}

\subsubsection{Analyses of Ginzburg--Landau coefficients}
Now let us move on to the main results, namely, the GL theory in the two-sublattice tetragonal model introduced in Sec.~\ref{sec:model_square2}.
Using Eqs.~\eqref{eq:GL_coefficient_alpha}, \eqref{eq:GL_coefficient_beta}, \eqref{eq:GL_coefficient_kappa}--\eqref{eq:GL_coefficient_delta'}, and \eqref{eq:GL_coefficient_alpha_tilde}, we computed the parameter dependence of the coefficients of the GL free energy, which determine the stability of different types of superconducting states.
We here assume a $d_{xy}$-wave order parameter,
\begin{equation}
 \hat{\phi}(\bm{k}) =
 \sqrt{2} \sin\frac{k_x}{2} \sin\frac{k_y}{2} \hat{\tau}_x \otimes \ii\hat{\sigma}_y,
 \label{eq:order_parameter_dxy}
\end{equation}
which belongs to the $B_2$ irreducible representation (irrep) of the point group $4mm$.
We confirmed that the $B_2$ order is the leading instability in many-body calculations of the onsite Hubbard model (Appendix~\ref{app:Eliashberg_Hubbard}).
For the calculations of Eqs.~\eqref{eq:GL_coefficient_alpha}--\eqref{eq:GL_coefficient_beta} and \eqref{eq:GL_coefficient_kappa}--\eqref{eq:GL_coefficient_delta'}, we use the analytic form of the sum in Matsubara frequencies, and numerically computed the sum in $\bm{k}$ using $512 \times 512$ meshes.

Figure~\ref{fig:phase_GL_square2} shows the zero lines of the GL coefficients in the parameter space of temperature ($T$) and altermagnetic molecular field ($h$).
As in the case of the uniform Zeeman field reviewed above, the superconducting transition temperature for small $h$ is determined by the $\alpha = 0$ line (red open squares), around which both $\beta$ and $\kappa$ are positive.
By increasing the altermagnetic field $h$, in contrast to the case of the uniform Zeeman field, along the $\alpha = 0$ line \textit{only} $\kappa$ changes its sign to negative (see the green dash-dotted line), whereas $\beta$ remains positive.
Therefore \textit{the phase-modulated FF state is stabilized in the high-altermagnetic-field regime}, according to Eqs.~\eqref{eq:free_energy_FF} and \eqref{eq:free_energy_LO}.
This is a distinctive feature in altermagnetic systems, in clear contrast to the amplitude-modulated LO state in the uniform Zeeman field.
We here note that the positive value of $\beta$ around the transition temperature is consistent with previous quasiclassical calculations, which reported the absence of first-order transitions in high exchange fields~\cite{Chourasia2024}.

\subsubsection{Mean-field theory}
To confirm the validity of our GL analysis, we evaluated the COM momentum of the Cooper pairs by self-consistent MF calculations.
Assuming the FF state $\eta(\bm{R}) = \eta_0 \ee^{2\ii\bm{Q} \cdot \bm{R}}$ ($\eta_0 \in \mathbb{R}$), we introduce the Bogoliubov--de Gennes Hamiltonian within the MF approximation by
\begin{align}
 H_{\text{MF}, \bm{Q}} &= \sum_{\bm{k}} \bm{\Psi}_{\bm{k}, \bm{Q}}^\dagger \hat{H}_{\bm{Q}}(\bm{k}) \bm{\Psi}_{\bm{k}, \bm{Q}} + \text{const.}, \\
 \hat{H}_{\bm{Q}}(\bm{k}) &=
 \begin{bmatrix}
  \hat{H}_0(\bm{k}_+) & \eta_0 \hat{\phi}(\bm{k}) \\
  \eta_0 \hat{\phi}(\bm{k})^\dagger & -\hat{H}_0(\bm{k}_-)^{\mathrm{T}}
 \end{bmatrix}, \\
 \bm{\Psi}_{\bm{k}, \bm{Q}}
 &= [c_{\bm{k}_+, a \uparrow}, c_{\bm{k}_+, a \downarrow}, c_{\bm{k}_+, b \uparrow}, c_{\bm{k}_+, b \downarrow}, \notag \\
 &\qquad c_{\bm{k}_-, a \uparrow}^\dagger, c_{\bm{k}_-, a \downarrow}^\dagger, c_{\bm{k}_-, b \uparrow}^\dagger, c_{\bm{k}_-, b \downarrow}^\dagger]^{\mathrm{T}}.
\end{align}
Then the grand potential in the superconducting state is given by
\begin{align}
 \Omega(\eta_0, \bm{Q}) = - \frac{T}{N} \sum_{\bm{k}} \tr\left[ \ln\left(1 + \ee^{-\hat{H}_{\bm{Q}}(\bm{k}) / T}\right) \right] + \frac{\eta_0^2}{V}.
\end{align}
The gap equation is represented as a derivative of the grand potential with respect to $\eta_0$:
\begin{equation}
 \frac{\partial}{\partial \eta_0} \Omega(\eta_0, \bm{Q}) = 0.
 \label{eq:gap_equation}
\end{equation}
Let $\eta_0 = \eta_0(\bm{Q})$ be the solution of the self-consistent Eq.~\eqref{eq:gap_equation}.
The condensation energy for the FF superconductivity is thus written as
\begin{equation}
 \mathcal{F}(\bm{Q}) = \Omega(\eta_0(\bm{Q}), \bm{Q}) - \Omega(0, \bm{Q}),
 \label{eq:condensation_energy}
\end{equation}
where $\Omega(0, \bm{Q}) = \Omega(0, \bm{0})$ holds.

\begin{figure*}
 \includegraphics[width=.65\linewidth, pagebox=artbox]{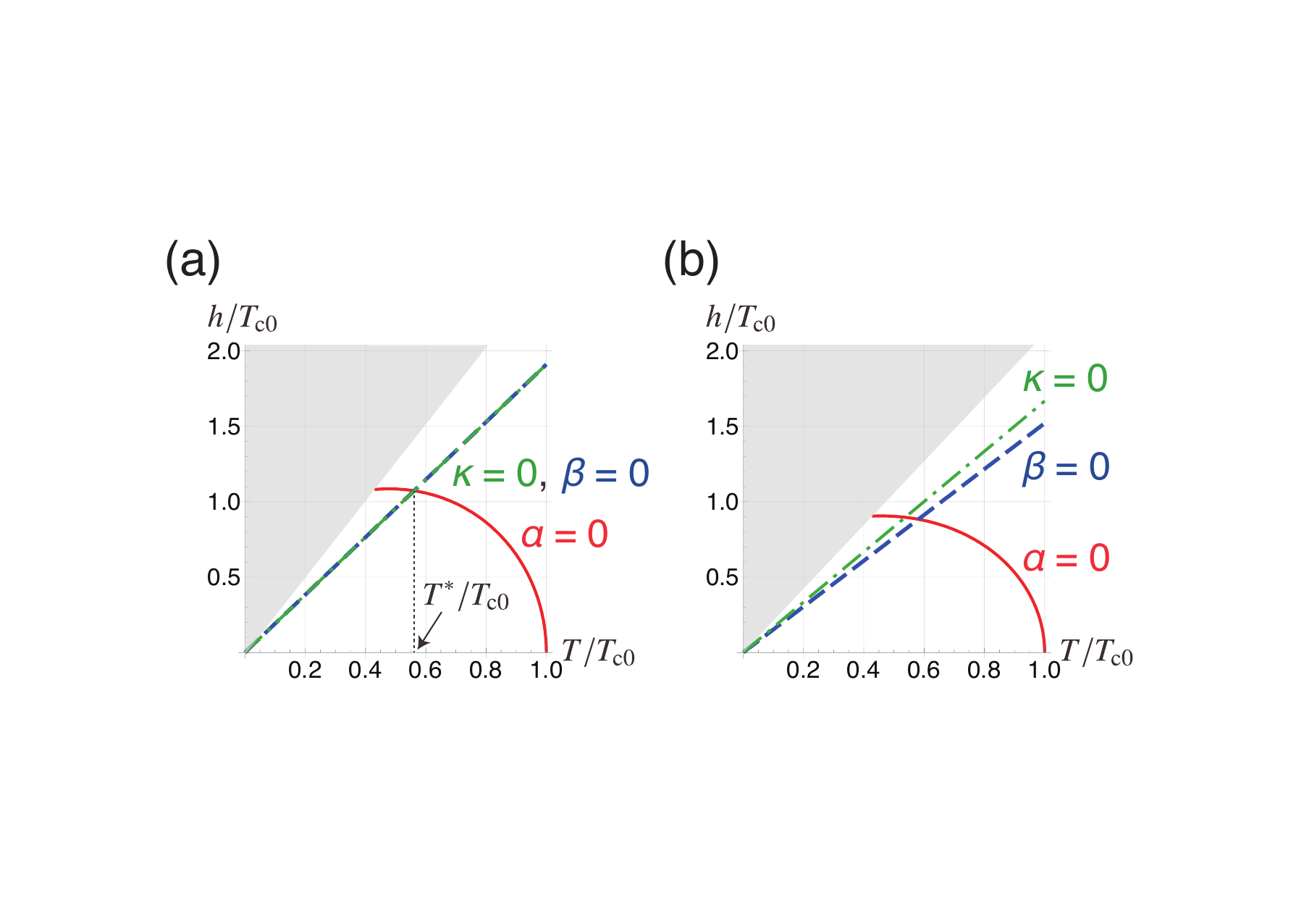}
 \caption{Temperature ($T$) vs field ($h$) phase diagrams for the continuum model for the (a) uniform and (b) $d$-wave Zeeman fields. Both axes are scaled by the transition temperature at zero magnetic field $T_{\mathrm{c}0}$. The $\alpha = 0$ lines are unreliable when $h / T$ is large (dark shaded areas) since we truncate the infinite series to the first 100 terms.}
 \label{fig:phase_GL_continuum_dS}
\end{figure*}

We numerically solve the gap Eq.~\eqref{eq:gap_equation}, and obtain the optimal COM momentum $\bm{Q}$ that minimizes the condensation energy \eqref{eq:condensation_energy}.
The results of the MF analysis are illustrated in Fig.~\ref{fig:phase_GL_square2}.
The filled (open) circles represent the superconducting (normal) state, in which the condensation energy has a negative (positive) sign.
The color of the filled circles indicates the magnitude of the optimal COM momentum $\bm{Q}$; the black circles signify $\bm{Q} = \bm{0}$ (BCS state), while the other colored ones represent the FF state with finite momentum.
Clearly, the FF state appears when the coefficient $\kappa$ of the gradient term is negative, and its transition temperature corresponds to the $\tilde{\alpha} = 0$ line.
Our calculations also indicate that $\bm{Q}$ is parallel to the $[11]$ direction.
This is consistent with our GL analysis, which shows that $\delta'$ is always negative near the FF phase and results in $\bm{Q} \parallel [11]$ [see Eq.~\eqref{eq:COM_momentum}].

\subsection{Continuum model}
\label{sec:GL_continuum}
We here investigate the GL coefficients in the continuum model introduced in Sec.~\ref{sec:model_continuum}.
In the following discussions, we assume $d$-wave order parameter $\hat{\phi}(\bm{k}_n) = \sqrt{2} (k_{nx}^2 - k_{ny}^2) \ii\hat{\sigma}_y$; however, the results are not qualitatively altered in $s$-wave superconductors.

\subsubsection{Uniform Zeeman field}
\label{sec:GL_continuum_mag_s}
First, we begin with the case in the uniform Zeeman field [Eq.~\eqref{eq:Hamiltonian_continuum_mag_s}], as in the original theory of the FFLO state under a magnetic field~\cite{Buzdin1997, Agterberg2001}.
Following the procedure in Refs.~\cite{Bauer_book, Wakatsuki2017, Wakatsuki2018}, we can analytically evaluate the GL coefficients in Eqs.~\eqref{eq:GL_coefficient_alpha}--\eqref{eq:GL_coefficient_kappa}.
Using the density of states $\rho_{\mathrm{F}}$ at the Fermi level, the coefficients are written as infinite series:
\begin{subequations}
 \label{eq:GL_continuum_sZ_dS}
 \begin{align}
  \alpha &= \frac{1}{V} - \rho_{\mathrm{F}} \sum_{n=0}^{\infty} (-1)^n S_{2n+1}(T) h^{2n},
  \label{eq:GL_continuum_sZ_dS_alpha} \displaybreak[2] \\
  \beta &= \frac{12m^2}{k_{\mathrm{F}}^2} \kappa
  = \frac{3\rho_{\mathrm{F}}}{2} \sum_{n=1}^{\infty} (-1)^{n-1} \binom{2n}{2} \, S_{2n+1}(T) h^{2n-2},
  \label{eq:GL_continuum_sZ_dS_beta_kappa}
 \end{align}
\end{subequations}
where we define
\begin{equation}
 S_l(T) := \pi T \sum_{\omega_n} \frac{1}{|\omega_n|^l} =
 \begin{cases}
  \displaystyle \log\frac{2 \ee^{\gamma_{\mathrm{E}}} E_{\mathrm{c}}}{\pi T} & l = 1, \\[2mm]
  \displaystyle \frac{(2^l - 1) \zeta(l)}{(2\pi T)^{l-1}} & l \geq 2,
 \end{cases}
\end{equation}
with $\gamma_{\mathrm{E}} \approx 0.577$ and $E_{\mathrm{c}}$ being Euler's constant and the energy cutoff, respectively.
For the derivation of Eqs.~\eqref{eq:GL_continuum_sZ_dS_alpha} and \eqref{eq:GL_continuum_sZ_dS_beta_kappa}, see Appendix~\ref{app:GL_coefficients_continuum}.
Importantly, Eq.~\eqref{eq:GL_continuum_sZ_dS_beta_kappa} shows that $\beta$ and $\kappa$ are exactly proportional to each other with a constant factor $12m^2 / k_{\mathrm{F}}^2$.
Therefore the two coefficients changes their sign simultaneously and the amplitude-modulated superconducting state, such as the LO state, is stabilized in a large magnetic field, as explained in Sec.~\ref{sec:GL_formalism}.
Taking the first 100 terms of the infinite series in Eqs.~\eqref{eq:GL_continuum_sZ_dS_alpha} and \eqref{eq:GL_continuum_sZ_dS_beta_kappa}, we compute the zeros of the GL coefficients, which are illustrated in Fig.~\ref{fig:phase_GL_continuum_dS}(a).
Note that, since the series are expanded by the powers of $h / T$, the calculation is not applicable to large $h / T$ shown as the shaded area.
Nevertheless, our calculations well reproduce the tricritical point $T = T^*$ at which all $\alpha$, $\beta$, and $\kappa$ becomes zero.
Indeed, this point is located at $T^* / T_{\mathrm{c}0} \approx 0.56$, which is consistent with previous studies~\cite{Saint-James_book, Shimahara1994, Shimahara1998, Matsuda_Shimahara_review, Buzdin1997, Yang1998, Agterberg2001}.

\subsubsection{\texorpdfstring{$d$-wave Zeeman field}{d-wave Zeeman field}}
\label{sec:GL_continuum_mag_d}
Next, let us consider the $d$-wave Zeeman field in Eq.~\eqref{eq:Hamiltonian_continuum_mag_d}.
The GL coefficients are given by
\begin{subequations}
 \label{eq:GL_continuum_dZ_dS}
 \begin{align}
  \alpha &= \frac{1}{V} - \rho_{\mathrm{F}} \sum_{n=0}^{\infty} (-1)^n \frac{(2n+2)!}{2^{n+1} [(n+1)!]^2} S_{2n+1}(T) h^{2n},
  \label{eq:GL_continuum_dZ_dS_alpha} \displaybreak[2] \\
  \beta &= \rho_{\mathrm{F}} \sum_{n=1}^{\infty} (-1)^{n-1} \frac{(2n)!}{2^n (n!)^2} n \cdot \frac{4n^2-1}{n+1} S_{2n+1}(T) h^{2n-2},
  \label{eq:GL_continuum_dZ_dS_beta} \displaybreak[2] \\
  \kappa &= \frac{\rho_{\mathrm{F}} k_{\mathrm{F}}^2}{8m^2} \sum_{n=1}^{\infty} (-1)^{n-1} \frac{(2n)!}{2^n (n!)^2} n \notag \\
  &\qquad \times \left[(2n-1) - \frac{2n+1}{n+1} \left(\frac{h}{E_{\mathrm{F}}}\right)^2\right] S_{2n+1}(T) h^{2n-2},
  \label{eq:GL_continuum_dZ_dS_kappa}
 \end{align}
\end{subequations}
where $E_{\mathrm{F}}$ is the Fermi energy; see Appendix~\ref{app:GL_coefficients_continuum} for the derivation.
Since superconductors typically satisfy the weak-coupling condition $T_{\mathrm{c}0} \ll E_{\mathrm{F}}$, we assume that the $d$-wave field is much smaller than the Fermi energy, namely $h \ll E_{\mathrm{F}}$, around the superconducting transition (for the analyses for strongly coupled superconductors, see Appendix~\ref{app:GL_coefficients_continuum_strong}).
Then the coefficient for the gradient term reduces to
\begin{equation}
 \kappa = \frac{\rho_{\mathrm{F}} k_{\mathrm{F}}^2}{8m^2} \sum_{n=1}^{\infty} (-1)^{n-1} \frac{(2n)!}{2^n (n!)^2} n (2n-1) S_{2n+1}(T) h^{2n-2},
 \label{eq:GL_continuum_dZ_dS_kappa_weak}
\end{equation}
where the summand is different from that in $\beta$ [Eq.~\eqref{eq:GL_continuum_dZ_dS_beta}], in contrast to the case of the uniform Zeeman field.
Figure~\ref{fig:phase_GL_continuum_dS}(b) shows the zeros of the GL coefficients in Eqs.~\eqref{eq:GL_continuum_dZ_dS_alpha}, \eqref{eq:GL_continuum_dZ_dS_beta}, and \eqref{eq:GL_continuum_dZ_dS_kappa_weak}.
Now, although the $\beta = 0$ and $\kappa = 0$ lines differ, we can see that the deviation is very small.
In this model, therefore, the LO state is expected to occur even in the $d$-wave Zeeman field.
This indicates that the FF state we obtained in Sec.~\ref{sec:GL_square2} cannot be explained by the Zeeman splitting in the continuum model.
This result is consistent with Ref.~\cite{Soto-Garrido2014}, which proposed the stability of the LO state and bidirectional PDW state in a triplet $\alpha$ nematic phase.

\subsection{Conventional square lattice model}
\label{sec:GL_square1}
Finally, let us apply the GL formalism to the conventional square lattice model introduced in Sec.~\ref{sec:model_square1}.
We here assume a $d_{x^2-y^2}$-wave order parameter,
\begin{equation}
 \hat{\phi}(\bm{k}') = (\cos k_x' - \cos k_y') \ii\hat{\sigma}_y,
\end{equation}
which corresponds to the $d_{xy}$-wave order in the two-sublattice tetragonal model [Eq.~\eqref{eq:order_parameter_dxy}]; recall that the coordinate axes in the $\bm{k}'$ space are rotated by $\pi/4$ radians from those in the $\bm{k}$ space (Fig.~\ref{fig:BZ_nodes}).

\begin{figure}
 \includegraphics[width=.8\linewidth, pagebox=artbox]{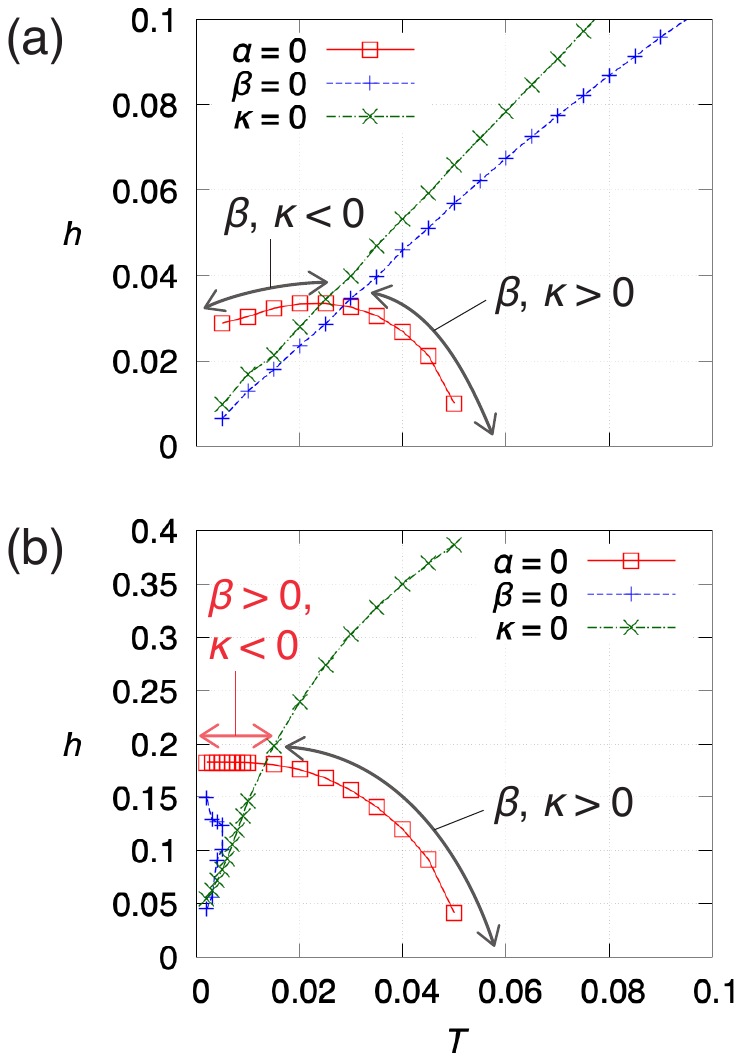}
 \caption{$T$--$h$ phase diagrams in the conventional square lattice model. We compute the GL coefficients in $1\bm{k}$-$d$-wave [Eq.~\eqref{eq:Hamiltonian_square1_mag_1}] and $2\bm{k}$-$d$-wave [\eqref{eq:Hamiltonian_square1_mag_2}] Zeeman fields for (a) and (b), respectively. The effective interaction is set to $V = 0.35$.}
 \label{fig:phase_GL_square1}
\end{figure}

Considering the $1\bm{k}$-$d$-wave Zeeman field [Eq.~\eqref{eq:Hamiltonian_square1_mag_1}], we numerically calculate the zeros of the GL coefficients [Eqs.~\eqref{eq:GL_coefficient_alpha}--\eqref{eq:GL_coefficient_kappa}], which are shown in Fig.~\ref{fig:phase_GL_square1}(a).
One can see that the coefficients $\beta$ and $\kappa$ flip their signs almost simultaneously, and the phase diagram resembles the case of continuum model with the $d$-wave Zeeman field in Fig.~\ref{fig:phase_GL_continuum_dS}(b).
Therefore, again, the LO state is stable in the high field regime, in contrast to the FF state stabilized in the two-sublattice tetragonal model.
We consider that the difference is caused by the form of the anisotropic Zeeman field, which does not reproduce the structure of the altermagnetic spin splitting in the two-sublattice tetragonal model, as mentioned in Sec.~\ref{sec:model_square1}.

Here, we propose an appropriate prescription of considering the $2\bm{k}$-$d$-wave Zeeman field [Eq.~\eqref{eq:Hamiltonian_square1_mag_2}], consistent with spin splitting in the two-sublattice tetragonal model, which in fact shows the stable altermagnetism-induced FF state.
Figure~\ref{fig:phase_GL_square1}(b) shows the GL phase diagram calculated in the $2\bm{k}$-$d$-wave Zeeman field.
In this case, the phase diagram resembles much to the two-sublattice tetragonal model in Fig.~\ref{fig:phase_GL_square2}, where only the coefficient $\kappa$ changes its sign when $\alpha \approx 0$, whereas $\beta$ is always positive around the transition temperature.
This means that the FF state can emerge.

Our analyses in this section comparing results for different models indicate that \textit{the multisublattice property is crucial for the occurrence of the altermagnetism-induced FF superconductivity}.
This conclusion is drawn from the stability of FF state over LO state in the cases of our two-sublattice tetragonal model and the conventional square lattice model having no sublattice degree of freedom but effectively accounting for the BZ folding in the two sublattice system by the $2\bm{k}$-$d$-wave Zeeman field.
We should note that, since we consider the carrier density $n = 0.85$ where the Fermi level is near the symmetry-protected coincident Van Hove singularity (VHS)~\cite{Yu2025} at the S point (see Fig.~\ref{fig:energy_band}), the difference of spin-split structure sensitively affects the properties near the Fermi surfaces and therefore crucially influences the resultant superconducting state.
Indeed, when we include the next-nearest-neighbor hoppings, which are neglected in both the two-sublattice tetragonal and conventional square lattice models, the phase-modulated superconductivity is found to be suppressed because the Fermi level shifts away from the coincident VHS.
Nevertheless, we confirmed that the FF state is restored by tuning the carrier density so that the Fermi level is close to the VHS.
While the coincident VHS is considered important for the altermagnetic instability~\cite{Yu2025}, how the singularity relates to the FF superconductivity remains an open question.

\section{Summary and Discussion}
\label{sec:summary}
In this study, we microscopically developed the GL theory in superconductors coexisting with an altermagnetic order, on the basis of three models: the two-sublattice tetragonal model, the continuum model, and the conventional square lattice model.
In the first model, as the altermagnetic molecular field becomes large, the GL coefficient $\kappa$ of the second-order gradient term changes sign from positive to negative, while the coefficient $\beta$ of the fourth-order uniform term remains positive around the superconducting transition temperature.
As a consequence, a phase-modulated FF superconductivity is stabilized.
This behavior is in drastic contrast with the well-known case of superconductivity in the uniform Zeeman field, where the amplitude-modulated LO state becomes stable due to the simultaneous sign change of both $\beta$ and $\kappa$.

We could identify the necessary ingredient for such exotic FF state to appear from the results of the other models:
The continuum model and the conventional square lattice model in a $1\bm{k}$-$d$-wave Zeeman field, which has often been used as the simplest representation of altermagnets in various studies~\cite{Zhang2024, Chakraborty2024, Chakraborty2024_arXiv_2, Sim2024, Hong2025, Mukasa2025, Iorsh2025, Hu2025_arXiv_1, Hu2025_arXiv_2}, exhibit the stability of amplitude-modulated superconductivity.
However, when we take into account the momentum doubling to the $d$-wave Zeeman field that reproduces the spin splitting structure in the two-sublattice tetragonal model, the FF state becomes stable than the LO state even in the conventional square lattice model.
Therefore the multisublattice degree of freedom is essential in realizing not only the altermagnetic state but also the exotic phase-modulated superconductivity.
This multisublattice effect differs from the so-called interband effect.
Indeed, we computed both intraband and interband parts of the GL coefficients in the two-sublattice tetragonal model and verified that the interband contribution is negligible compared to the intraband component.

Here we comment on the relevance of our theory to previous studies.
The phase-modulated superconductivity with a single $\bm{Q}$ has been proposed in the context of noncentrosymmetric superconductors~\cite{Edelstein1989, Mineev1994, Agterberg2003, Dimitrova2003, Samokhin2004, Kaur2005, Agterberg2007, Dimitrova2007, Mineev2008, Smidman2017_review}, odd-parity magnetic multipoles~\cite{Sumita2016, Sumita2017, Amin2024}, and anapole superconductivity~\cite{Kanasugi2022, Kitamura2023} (in the first context, it is also called as helical superconductivity).
In addition, a recent theoretical study has suggested the emergence of an FF state in valley-polarized superconductors, which have a Fermi surface similar to that in $f$-wave altermagnets~\cite{Daido2025_arXiv}.
All the phase-modulated states discussed in these previous studies are attributed to odd-order terms concerning spatial derivatives, which can appear in the GL free energy by breaking the inversion and time-reversal symmetries~%
\footnote{In particular, terms with the second order in $\eta$ and \textit{linear} in spatial gradient are called the Lifshitz invariants.}.
In contrast, in our case the FF state can be stabilized even in the absence of odd orders; this situation is rather reminiscent of impurity effects under a uniform magnetic field~\cite{Agterberg2001, Houzet2006}.

For the detection of the phase modulation in FF (helical) superconductivity, an interference experiment based on the Josephson effect has been proposed~\cite{Yang2000, Kaur2005, Kim2016}.
We believe that this method will also be effective in observing the FF state in altermagnets.
We have studied clean superconducting systems without spin--orbit coupling, while the conventional FFLO superconducting state under a uniform magnetic field is known to be fragile against disorder, thermal fluctuations, or spin--orbit coupling.
The stability of the altermagnetism-induced FF state against such perturbations is crucial for its experimental detection, whose investigations are left for future work.

Our study shows that altermagnets constitute a novel and effective platform for phase-modulated FF superconductivity in the absence of an external magnetic field.
A key ingredient for this exotic state is the multisublattice degree of freedom.
As a future direction, cluster multipole systems, which includes altermagnets, are unexplored and interesting platforms to realize exotic quantum phenomena.

\begin{acknowledgments}
 The authors are grateful to Taira Kawamura, Yusuke Kato, Akito Daido, and Taisei Kitamura for valuable comments.
 This work was supported by Japan Society for the Promotion of Science (JSPS) KAKENHI Grants No. JP25H02113, No. JP23K03333, No. JP23K13056, No. JP23H01129, and No. JP23H04047.
 Numerical calculations in this work were partly performed using the facilities of the Supercomputer Center, the Institute for Solid State Physics, the University of Tokyo.
\end{acknowledgments}

\appendix

\begin{figure*}
 \includegraphics[width=.65\linewidth, pagebox=artbox]{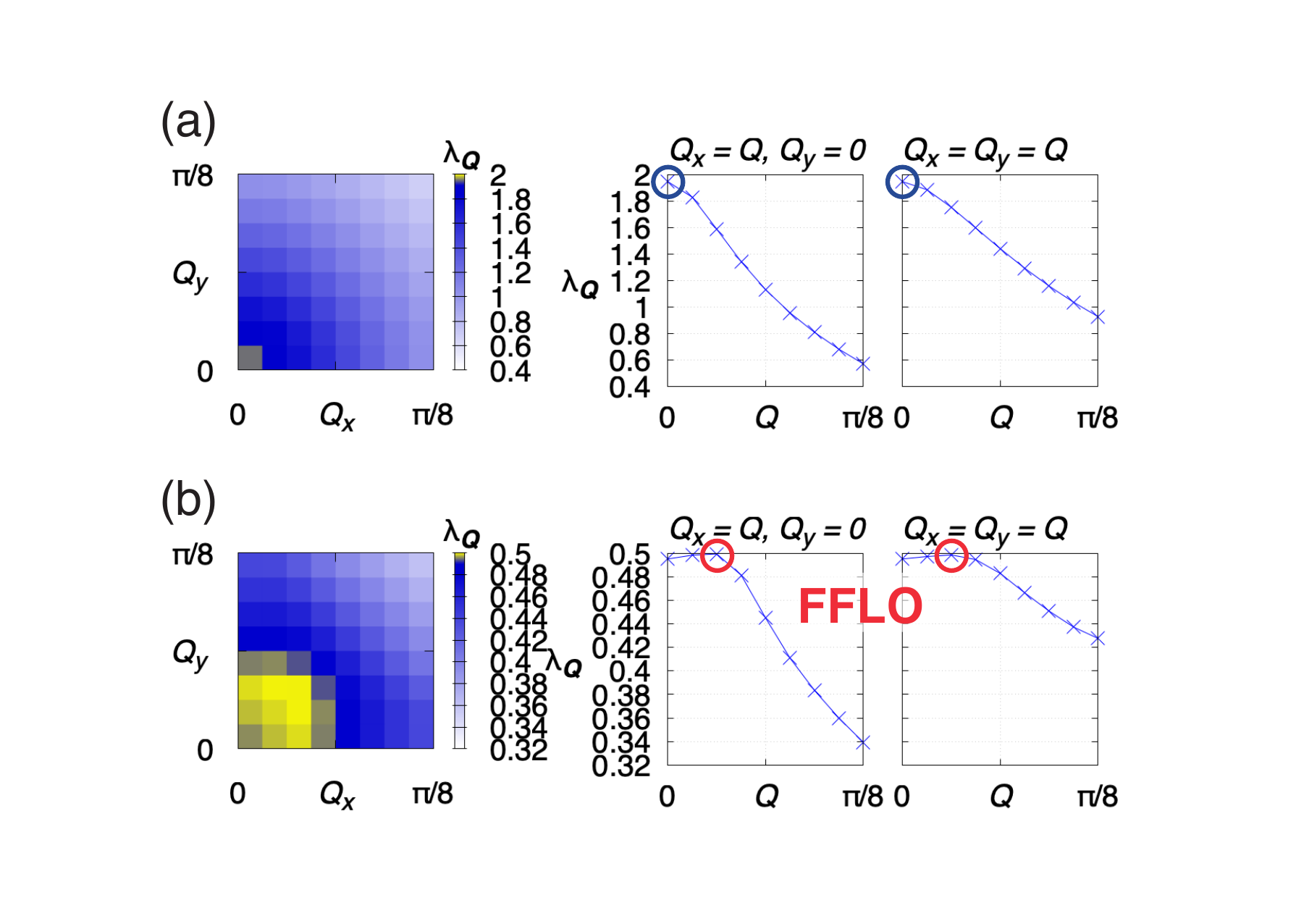}
 \caption{The COM momentum dependence of the eigenvalue $\lambda_{\bm{Q}}$ of the Eliashberg Eq.~\eqref{eq:linearized_Eliashberg} for (a) $h = 0$ and (b) $h = 0.4$. The eigenvalue has peaks at finite momenta in (b).}
 \label{fig:B2_lam_h001_a_0.000}
\end{figure*}

\section{Eliashberg theory based on Hubbard model}
\label{app:Eliashberg_Hubbard}
For the analyses of the two-sublattice tetragonal model in the main text, we assume the $d_{xy}$-wave superconducting order belonging to the $B_2$ irrep [Eq.~\eqref{eq:order_parameter_dxy}] within the MF approximation.
Here, we demonstrate that this $d_{xy}$-wave superconducting state is indeed the leading instability and the FFLO state can be stabilized on the basis of the Hubbard interaction and Eliashberg theory.

\subsection{Formalism}
In the following, we discuss the Hubbard Hamiltonian $H = H_0 + H_{\text{int}}^{\text{(Hub)}}$, where the interaction term is given by
\begin{equation}
 H_{\text{int}}^{\text{(Hub)}} = U \sum_{\bm{R}} \sum_{l} n_{\bm{R}, l\uparrow} n_{\bm{R}, l\downarrow},
 \label{eq:interaction_Hubbard}
\end{equation}
with $U > 0$ being the onsite Coulomb repulsion.
$n_{\bm{R}, l s} = c_{\bm{R}, l s}^\dagger c_{\bm{R}, l s}$ is the electron-number operator.

To consider fluctuations of electric and magnetic multipoles, let us formulate the generalized susceptibility within the random-phase approximation.
This is calculated as
\begin{equation}
 \hat{\chi}(\bm{q}; \Omega_m) = \left[ \hat{\bm{1}} - \hat{\chi}_0(\bm{q}; \Omega_m) \hat{U}^{\text{(Hub)}} \right]^{-1} \hat{\chi}_0(\bm{q}; \Omega_m),
 \label{eq:generalized_susceptibility}
\end{equation}
where $\Omega_m = 2m\pi T$ is the bosonic Matsubara frequency.
The bare susceptibility $\chi_0$ is defined by
\begin{equation}
 \chi_0(q)_{\zeta_1 \zeta_2, \zeta_3 \zeta_4} = - \frac{T}{N} \sum_{k} G_0(k+q)_{\zeta_1, \zeta_3} G_0(k)_{\zeta_4, \zeta_2},
\end{equation}
where we use the abbreviations $k = (\bm{k}; \omega_n)$ and $q = (\bm{q}; \Omega_m)$.
The matrix elements of the Hubbard interaction $\hat{U}^{\text{(Hub)}}$ are defined by
\begin{gather}
 U^{\text{(Hub)}}_{\zeta_1 \zeta_2, \zeta_3 \zeta_4} = \delta_{l_1, l_2} \delta_{l_2, l_3} \delta_{l_3, l_4} U_{s_1 s_2, s_3 s_4}, \\
 U_{s_1 s_2, s_3 s_4} =
 \begin{cases}
  U & (s_1 s_2, s_3 s_4) = (\uparrow \downarrow, \uparrow \downarrow) \ \text{or} \  (\downarrow \uparrow, \downarrow \uparrow), \\
  -U & (s_1 s_2, s_3 s_4) = (\uparrow \uparrow, \downarrow \downarrow) \ \text{or} \  (\downarrow \downarrow, \uparrow \uparrow), \\
  0 & \text{otherwise}.
 \end{cases}
\end{gather}

Next, we construct the linearized Eliashberg equation with finite COM momentum~\cite{Yoshida2021, Sumita2023}:
\begin{equation}
 \lambda_{\bm{Q}} \Delta_{\bm{Q}}(k)_{\zeta, \zeta'} = \frac{T}{N} \sum_{q} \sum_{\zeta_1, \zeta_2} V^{\text{(a)}}(q)_{\zeta \zeta_1, \zeta_2 \zeta'} F_{\bm{Q}}(k-q)_{\zeta_1, \zeta_2},
 \label{eq:linearized_Eliashberg}
\end{equation}
where $\hat{\Delta}_{\bm{Q}}(k)$ is the order parameter matrix, and the anomalous Green's function is given by
\begin{equation}
 \hat{F}_{\bm{Q}}(k) = - \hat{G}_0(k) \hat{\Delta}_{\bm{Q}}(k) \hat{G}_0(-k+Q)^{\mathrm{T}}, 
 \label{eq:Green_func_anomalous}
\end{equation}
with $Q = (\bm{Q}; 0)$.
Since we consider the multipole fluctuations as a glue of pairing, the interaction vertex are represented by using the generalized susceptibility \eqref{eq:generalized_susceptibility}:
\begin{align}
 \hat{V}^{\text{(a)}}(q) &= - \frac{1}{2} \hat{U}^{\text{(Hub)}} - \hat{U}^{\text{(Hub)}} \hat{\chi}(q) \hat{U}^{\text{(Hub)}}.
\end{align}
In this finite-momentum Eliashberg theory, we can discuss the instability of the FF state with the COM momentum $\bm{Q}$ for a given momentum $\bm{Q}$.
The phase transition into the FF state takes place when the eigenvalue $\lambda_{\bm{Q}}$ in Eq.~\eqref{eq:linearized_Eliashberg} reaches unity.

\subsection{Numerical results}
Using the above formalism, let us investigate the possible superconducting order in an altermagnetic state.
By using the power method, we solve the linearized Eliashberg Eq.~\eqref{eq:linearized_Eliashberg}.
For the practical reason, we fix the temperature and seek the optimal combination of the COM momentum $\bm{Q}$ and order parameter $\hat{\Delta}_{\bm{Q}}(k)$ that provides the largest eigenvalue $\lambda_{\bm{Q}}$.
In the numerical study, we take $128 \times 128$ $\bm{k}$-point meshes and about 90 Matsubara frequencies generated by \texttt{SparseIR.jl} package~\cite{Wallerberger2023, Wang2020} based on the intermediate representation~\cite{Shinaoka2017} and the sparse sampling~\cite{Li2020}.
The temperature $T$ and the energy cutoff $\omega_{\text{max}}$ are set to $0.01t_1$ and $32t_1$, respectively.
We consider the SC order parameter belonging to the four irreps of $4mm$: $A_1$, $A_2$, $B_1$, and $B_2$.
Using $\tilde{\bm{k}} := \bm{k} - \bm{Q}/2$, the initial functional form is chosen as
\begin{equation}
 \hat{\Delta}_{\bm{Q}}^{\text{(init)}}(k) \sim
 \begin{cases}
  \hat{\tau}_0 \otimes \ii\hat{\sigma}_y & \text{for $A_1$}, \\
  (\cos\tilde{k}_x - \cos\tilde{k}_y) (\hat{\tau}_z \otimes \ii\hat{\sigma}_y) & \text{for $A_2$}, \\
  (\cos\tilde{k}_x - \cos\tilde{k}_y) (\hat{\tau}_0 \otimes \ii\hat{\sigma}_y) & \text{for $B_1$}, \\
  \sin\frac{\tilde{k}_x}{2} \sin\frac{\tilde{k}_y}{2} (\hat{\tau}_x \otimes \ii\hat{\sigma}_y) & \text{for $B_2$}.
 \end{cases}
\end{equation}
We confirmed that the $B_2$ representation ($d_{xy}$-wave superconductivity) is the leading instability for superconductivity in our model for $0 \leq h \leq 0.4$.

Figures~\ref{fig:B2_lam_h001_a_0.000} shows the $\bm{Q}$ dependence of the Eliashberg eigenvalue.
In the paramagnetic state ($h = 0$), the eigenvalue has a single peak at zero momentum [Fig.~\ref{fig:B2_lam_h001_a_0.000}(a)].
On the other hand, the maximum is located at finite $\bm{Q}$ in the altermagnetic state with $h = 0.4$ [Fig.~\ref{fig:B2_lam_h001_a_0.000}(b)].
This results indicate that the FFLO superconductivity is stabilized under the altermagnetic order.

\section{Microscopic derivation of GL equation}
\label{app:GL_microscopic_derivation}
We show the detailed derivation of the GL Eq.~\eqref{eq:GL} from the Gor'kov equations~\eqref{eq:Gorkov}.
Here, we assume that the vector potential is absent, neglecting the orbital pair breaking effect due to an external magnetic field~%
\footnote{When the vector potential $\bm{A}(\bm{R})$ is taken into account, one should replace the spatial derivative $\nabla_{\bm{R}}$ with the covariant derivative $\bm{D}_{\bm{R}} := \nabla_{\bm{R}} - \ii \frac{2e}{\hbar c} \bm{A}(\bm{R})$ in the final expression.}.
Therefore the normal Green's function of free electrons is given by
\begin{align}
 \hat{G}_0(\bm{x}, \bm{x}'; \omega_n)
 &= \hat{G}_0(\bm{r}; \omega_n) \notag \\
 &= \frac{1}{N} \sum_{\bm{k}} \ee^{\ii \bm{k} \cdot \bm{r}} \left[(\ii\omega_n + \mu) \hat{\bm{1}} - \hat{H}_0(\bm{k})\right]^{-1},
 \label{eq:G0_r}
\end{align}
which is a function of the relative coordinate $\bm{r} := \bm{x} - \bm{x}'$ because of the translation symmetry.
Using the Wigner representation of the order parameter [Eq.~\eqref{eq:order_parameter_Wigner}], the Gor'kov equations~\eqref{eq:Gorkov} are rewritten as
\begin{widetext}
\begin{subequations}
 \label{eq:Gorkov_2}
 \begin{align}
  \hat{G}\left(\bm{x}, \bm{x}'; \omega_n\right)
  &= \hat{G}_0\left(\bm{r}; \omega_n\right)
  - \frac{1}{N} \sum_{\bm{r}_{12}, \bm{R}_{12}} \sum_{\bm{k}'} \eta(\bm{R}_{12}) \ee^{\ii \bm{k}' \cdot \bm{r}_{12}} \hat{G}_0\left(\bm{R} + \frac{\bm{r}}{2} - \bm{R}_{12} - \frac{\bm{r}_{12}}{2}; \omega_n\right) \hat{\phi}(\bm{k}') \hat{F}\left(\bm{R}_{12} - \frac{\bm{r}_{12}}{2}, \bm{R} - \frac{\bm{r}}{2}; \omega_n\right)^*, \displaybreak[2] \\
  \hat{F}\left(\bm{x}, \bm{x}'; \omega_n\right)
  &= - \frac{1}{N} \sum_{\bm{r}_{12}, \bm{R}_{12}} \sum_{\bm{k}'} \eta(\bm{R}_{12}) \ee^{\ii \bm{k}' \cdot \bm{r}_{12}} \hat{G}_0\left(\bm{R} + \frac{\bm{r}}{2} - \bm{R}_{12} - \frac{\bm{r}_{12}}{2}; \omega_n\right) \hat{\phi}(\bm{k}') \hat{G}\left(\bm{R}_{12} - \frac{\bm{r}_{12}}{2}, \bm{R} - \frac{\bm{r}}{2}; \omega_n\right)^*,
 \end{align}
\end{subequations}
where we define $\bm{R} := (\bm{x} + \bm{x}') / 2$, $\bm{r}_{12} := \bm{x}_1 - \bm{x}_2$, and $\bm{R}_{12} := (\bm{x}_1 + \bm{x}_2) / 2$.
Using the above equations, we expand the right-hand side (RHS) of the MF Eq.~\eqref{eq:mean-field} up to the third order of $\eta$:
\begin{align}
 \text{[RHS of Eq.~\eqref{eq:mean-field}]}
 &= - \frac{TV}{N} \sum_{\omega_n} \sum_{\bm{r}} \sum_{\bm{k}} \ee^{-\ii \bm{k} \cdot \bm{r}} \tr\left[
  \hat{\phi}(\bm{k})^\dagger \hat{F}\left(\bm{R} + \frac{\bm{r}}{2}, \bm{R} - \frac{\bm{r}}{2}; \omega_n\right) 
 \right] \notag \displaybreak[2] \\
 &= \frac{TV}{N^2} \sum_{\omega_n} \sum_{\bm{r}} \sum_{\bm{r}_{12}, \bm{R}_{12}} \sum_{\bm{k}, \bm{k}'} \eta(\bm{R}_{12}) \ee^{-\ii \bm{k} \cdot \bm{r} + \ii \bm{k}' \cdot \bm{r}_{12}} \notag \\
 &\qquad \times \tr\left[
  \hat{\phi}(\bm{k})^\dagger \hat{G}_0\left(\bm{R} + \frac{\bm{r}}{2} - \bm{R}_{12} - \frac{\bm{r}_{12}}{2}; \omega_n\right)
  \hat{\phi}(\bm{k}') \hat{G}\left(\bm{R}_{12} - \frac{\bm{r}_{12}}{2}, \bm{R} - \frac{\bm{r}}{2}; \omega_n\right)^*
 \right] \notag \displaybreak[2] \\
 &= \frac{TV}{N^2} \sum_{\omega_n} \sum_{\bm{r}} \sum_{\bm{r}_{12}, \bm{R}_{12}} \sum_{\bm{k}, \bm{k}'} \eta(\bm{R}_{12}) \ee^{-\ii \bm{k} \cdot \bm{r} + \ii \bm{k}' \cdot \bm{r}_{12}} \notag \\
 &\qquad \times \tr\left[
  \hat{\phi}(\bm{k})^\dagger \hat{G}_0\left(\bm{R} + \frac{\bm{r}}{2} - \bm{R}_{12} - \frac{\bm{r}_{12}}{2}; \omega_n\right)
  \hat{\phi}(\bm{k}') \hat{G}_0\left(\bm{R}_{12} - \frac{\bm{r}_{12}}{2} - \bm{R} + \frac{\bm{r}}{2}; \omega_n\right)^*
 \right] \notag \\
 &\quad - \frac{TV}{N^3} \sum_{\omega_n} \sum_{\bm{r}} \sum_{\bm{r}_{12}, \bm{R}_{12}} \sum_{\bm{r}_{34}, \bm{R}_{34}} \sum_{\bm{k}, \bm{k}', \bm{k}''} \eta(\bm{R}_{12}) \eta(\bm{R}_{34})^* \ee^{-\ii \bm{k} \cdot \bm{r} + \ii \bm{k}' \cdot \bm{r}_{12} - \ii \bm{k}'' \cdot \bm{r}_{34}} \notag \\
 &\qquad \times \tr\left[
  \hat{\phi}(\bm{k})^\dagger \hat{G}_0\left(\bm{R} + \frac{\bm{r}}{2} - \bm{R}_{12} - \frac{\bm{r}_{12}}{2}; \omega_n\right)
  \hat{\phi}(\bm{k}') \hat{G}_0\left(\bm{R}_{12} - \frac{\bm{r}_{12}}{2} - \bm{R}_{34} - \frac{\bm{r}_{34}}{2}; \omega_n\right)^*
 \right. \notag \\
 &\qquad\qquad \times
 \left.
  \hat{\phi}(\bm{k}'')^* \hat{F}\left(\bm{R}_{34} - \frac{\bm{r}_{34}}{2}, \bm{R} - \frac{\bm{r}}{2}; \omega_n\right)
 \right] \notag \displaybreak[2] \\
 &\simeq \frac{TV}{N^2} \sum_{\omega_n} \sum_{\bm{r}} \sum_{\bm{r}_{12}, \bm{R}_{12}} \sum_{\bm{k}, \bm{k}'} \eta(\bm{R}_{12}) \ee^{-\ii \bm{k} \cdot \bm{r} + \ii \bm{k}' \cdot \bm{r}_{12}} \notag \\
 &\qquad \times \tr\left[
  \hat{\phi}(\bm{k})^\dagger \hat{G}_0\left(\bm{R} + \frac{\bm{r}}{2} - \bm{R}_{12} - \frac{\bm{r}_{12}}{2}; \omega_n\right)
  \hat{\phi}(\bm{k}') \hat{G}_0\left(\bm{R}_{12} - \frac{\bm{r}_{12}}{2} - \bm{R} + \frac{\bm{r}}{2}; \omega_n\right)^*
 \right] \notag \\
 &\quad - \frac{TV}{N^4} \sum_{\omega_n} \sum_{\bm{r}} \sum_{\bm{r}_{12}, \bm{R}_{12}} \sum_{\bm{r}_{34}, \bm{R}_{34}} \sum_{\bm{r}_{56}, \bm{R}_{56}} \sum_{\bm{k}, \bm{k}', \bm{k}'', \bm{k}'''} \eta(\bm{R}_{12}) \eta(\bm{R}_{34})^* \eta(\bm{R}_{56}) \ee^{-\ii \bm{k} \cdot \bm{r} + \ii \bm{k}' \cdot \bm{r}_{12} + \ii \bm{k}'' \cdot \bm{r}_{34} + \ii \bm{k}''' \cdot \bm{r}_{56}} \notag \\
 &\qquad \times \tr\left[
  \hat{\phi}(\bm{k})^\dagger \hat{G}_0\left(\bm{R} + \frac{\bm{r}}{2} - \bm{R}_{12} - \frac{\bm{r}_{12}}{2}; \omega_n\right)
  \hat{\phi}(\bm{k}') \hat{G}_0\left(\bm{R}_{12} - \frac{\bm{r}_{12}}{2} - \bm{R}_{34} - \frac{\bm{r}_{34}}{2}; \omega_n\right)^*
 \right. \notag \\
 &\qquad\qquad \times
 \left.
  \hat{\phi}(\bm{k}'')^\dagger \hat{G}_0\left(\bm{R}_{34} - \frac{\bm{r}_{34}}{2} - \bm{R}_{56} - \frac{\bm{r}_{56}}{2}; \omega_n\right)
  \hat{\phi}(\bm{k}''') \hat{G}_0\left(\bm{R}_{56} - \frac{\bm{r}_{56}}{2} - \bm{R} + \frac{\bm{r}}{2}; \omega_n\right)^*
 \right].
 \label{eq:MF_RHS}
\end{align}
In the second term of the final expression, we use the transformation $\bm{k}'' \to -\bm{k}''$ and replace $\hat{G}$ with $\hat{G}_0$ as an approximation.
Using Eq.~\eqref{eq:G0_r}, the first term in Eq.~\eqref{eq:MF_RHS} is further calculated as
\begin{align}
 \text{[The first term in Eq.~\eqref{eq:MF_RHS}]}
 &= \frac{TV}{N^2} \sum_{\omega_n} \sum_{\bm{r}} \sum_{\bm{r}_{12}, \bm{R}_{12}} \sum_{\bm{k}, \bm{k}'} \eta(\bm{R}_{12}) \ee^{-\ii \bm{k} \cdot \bm{r} + \ii \bm{k}' \cdot \bm{r}_{12}} \notag \\
 &\qquad \times \frac{1}{N^2} \sum_{\bm{k}_1, \bm{k}_2} \ee^{\ii \bm{k}_1 \cdot (\bm{R} + \frac{\bm{r}}{2} - \bm{R}_{12} - \frac{\bm{r}_{12}}{2})} \ee^{-\ii \bm{k}_2 \cdot (\bm{R}_{12} - \frac{\bm{r}_{12}}{2} - \bm{R} + \frac{\bm{r}}{2})}
 \tr\left[
  \hat{\phi}(\bm{k})^\dagger \hat{G}_0\left(\bm{k}_1; \omega_n\right)
  \hat{\phi}(\bm{k}') \hat{G}_0\left(\bm{k}_2; \omega_n\right)^*
 \right] \notag \\
 &= \frac{TV}{N^2} \sum_{\omega_n} \sum_{\bm{k}, \bm{Q}} \sum_{\bm{R}_{12}}
 \underline{\ee^{-2\ii \bm{Q} \cdot (\bm{R}_{12} - \bm{R})} \eta(\bm{R}_{12})}
 \tr\left[ \hat{X}(\bm{k}, \bm{Q}; \omega_n) \right].
\end{align}
Defining $\bar{\bm{R}} := \bm{R}_{12} - \bm{R}$, we expand $\eta(\bm{R}_{12}) = \eta(\bm{R} + \bar{\bm{R}})$ around $\bm{R}$.
Then the underlined part is given by
\begin{align}
 \ee^{-2\ii \bm{Q} \cdot \bar{\bm{R}}} \eta(\bm{R} + \bar{\bm{R}})
 &= \ee^{-2\ii \bm{Q} \cdot \bar{\bm{R}}}
 \sum_{n=0}^{\infty} \frac{1}{n!} (\bar{\bm{R}} \cdot \nabla_{\bm{R}'})^n \eta(\bm{R}')\bigg|_{\bm{R}' = \bm{R}} \notag \\
 &= \ee^{-2\ii \bm{Q} \cdot \bar{\bm{R}}} \eta(\bm{R})
 + \sum_{n=1}^{\infty} \frac{1}{n!} \left(\frac{-1}{2\ii}\right)^n \sum_{\mu_1, \dots, \mu_n} \frac{\partial^n \ee^{-2\ii \bm{Q} \cdot \bar{\bm{R}}}}{\partial Q_{\mu_1} \dotsb \partial Q_{\mu_n}}
 \cdot \frac{\partial^n \eta(\bm{R})}{\partial R_{\mu_1} \dotsb \partial R_{\mu_n}}.
\end{align}
Assuming the slowly varying of the order parameter in the real space, we take the expansion up to the fourth order with respect to the derivative $\nabla_{\bm{R}}$.
Then we obtain
\begin{align}
 \text{[The first term in Eq.~\eqref{eq:MF_RHS}]}
 &\simeq \frac{TV}{N^2} \eta(\bm{R}) \sum_{\omega_n} \sum_{\bm{k}, \bm{Q}}
 \sum_{\bar{\bm{R}}} \ee^{-2\ii \bm{Q} \cdot \bar{\bm{R}}}
 \tr\left[ \hat{X}(\bm{k}, \bm{Q}; \omega_n) \right] \notag \\
 &\quad + \frac{TV}{N^2} \sum_{n=1}^{4} \frac{1}{n!} \left(\frac{-1}{2\ii}\right)^n \sum_{\mu_1, \dots, \mu_n}
 \frac{\partial^n \eta(\bm{R})}{\partial R_{\mu_1} \dotsb \partial R_{\mu_n}}
 \sum_{\omega_n} \sum_{\bm{k}, \bm{Q}} \sum_{\bar{\bm{R}}}
 \frac{\partial^n \ee^{-2\ii \bm{Q} \cdot \bar{\bm{R}}}}{\partial Q_{\mu_1} \dotsb \partial Q_{\mu_n}}
 \cdot \tr\left[ \hat{X}(\bm{k}, \bm{Q}; \omega_n) \right] \notag \displaybreak[2] \\
 &= \frac{TV}{N} \eta(\bm{R}) \sum_{\omega_n} \sum_{\bm{k}}
 \tr\left[ \hat{X}(\bm{k}, \bm{0}; \omega_n) \right] \notag \\
 &\quad + \frac{TV}{N} \sum_{n=1}^{4} \frac{1}{n! (2\ii)^n} \sum_{\mu_1, \dots, \mu_n}
 \frac{\partial^n \eta(\bm{R})}{\partial R_{\mu_1} \dotsb \partial R_{\mu_n}}
 \sum_{\omega_n} \sum_{\bm{k}} \frac{\partial^n}{\partial Q_{\mu_1} \dotsb \partial Q_{\mu_n}}
 \tr\left[ \hat{X}(\bm{k}, \bm{Q}; \omega_n) \right] \bigg|_{\bm{Q} = \bm{0}} \notag \\
 &= (1 - V \alpha) \eta(\bm{R})
 - V \sum_{n=1}^{4} \sum_{\mu_1, \dots, \mu_n} c^{(2)}_{\mu_1 \dots \mu_n}
 \frac{\partial^n \eta(\bm{R})}{\partial R_{\mu_1} \dotsb \partial R_{\mu_n}},
 \label{eq:Gorkov_RHS_1st}
\end{align}
where we use Eqs.~\eqref{eq:GL_coefficient_alpha} and \eqref{eq:c2_diff}.
On the other hand, the third-order term in $\eta$ is given by
\begin{equation}
 \text{[The second term in Eq.~\eqref{eq:MF_RHS}]}
 \simeq - \frac{TV}{N} |\eta(\bm{R})|^2 \eta(\bm{R}) \sum_{\omega_n} \sum_{\bm{k}}
 \tr\left[ \hat{X}(\bm{k}, \bm{0}; \omega_n)^2 \right]
 = - 2V \beta |\eta(\bm{R})|^2 \eta(\bm{R}),
 \label{eq:Gorkov_RHS_2nd}
\end{equation}
where we neglect the spatial derivative and use Eq.~\eqref{eq:GL_coefficient_beta}.
Substituting Eqs.~\eqref{eq:Gorkov_RHS_1st} and \eqref{eq:Gorkov_RHS_2nd} into the MF Eq.~\eqref{eq:mean-field}, we finally get the GL equation:
\begin{gather}
 \eta(\bm{R}) = (1 - V \alpha) \eta(\bm{R}) - 2V \beta |\eta(\bm{R})|^2 \eta(\bm{R})
 - V \sum_{n=1}^{4} \sum_{\mu_1, \dots, \mu_n} c^{(2)}_{\mu_1 \dots \mu_n}
 \frac{\partial^n \eta(\bm{R})}{\partial R_{\mu_1} \dotsb \partial R_{\mu_n}}, \notag \\
 \therefore \alpha \eta(\bm{R}) + 2\beta |\eta(\bm{R})|^2 \eta(\bm{R})
 + \sum_{n=1}^{4} \sum_{\mu_1, \dots, \mu_n} c^{(2)}_{\mu_1 \dots \mu_n}
 \frac{\partial^n \eta(\bm{R})}{\partial R_{\mu_1} \dotsb \partial R_{\mu_n}} = 0.
\end{gather}
Imposing the symmetry of tetragonal systems on this equation, we can reproduce Eq.~\eqref{eq:GL} in the main text.

\section{GL coefficients in a continuum model}
\label{app:GL_coefficients_continuum}
Following Refs.~\cite{Bauer_book, Wakatsuki2017, Wakatsuki2018}, we derive the GL coefficients of a continuum model in a Zeeman field [Eqs.~\eqref{eq:GL_continuum_sZ_dS} and \eqref{eq:GL_continuum_dZ_dS}].
For the Hamiltonian in Eqs.~\eqref{eq:Hamiltonian_continuum_kin} and \eqref{eq:Hamiltonian_continuum_mag}, the Green's function of free electrons is given by
\begin{equation}
 G_0(\bm{k}; \omega_n)_{s, s'} = \frac{\delta_{s, s'}}{\ii\omega_n - \xi(\bm{k}) + (-1)^s h(\bm{k})},
\end{equation}
where $\xi(\bm{k}) = \frac{\bm{k}^2}{2m} - \mu$.
Using this Green's function, we calculate the coefficients $\alpha$, $\beta$, and $\kappa$ [Eqs.~\eqref{eq:GL_coefficient_alpha}--\eqref{eq:GL_coefficient_beta} and \eqref{eq:GL_coefficient_kappa}] by replacing the summation $\frac{1}{N} \sum_{\bm{k}}$ with an integral $\frac{1}{(2\pi)^2} \int \mathrm{d}\bm{k}$.
For the analytical calculations in the following, we will refer to some useful formulas, which are summarized at the end of this Appendix (Sec.~\ref{app:GL_coefficients_continuum_formulas}).

First, to discuss $\alpha$ and $\kappa$, let us consider the following integral,
\begin{equation}
 I(\bm{Q}) := T \sum_{\omega_n} \int \frac{\mathrm{d}\bm{k}}{(2\pi)^2} \tr\left[\hat{X}(\bm{k}, \bm{Q}; \omega_n)\right],
\end{equation}
whose expansion with respect to $\bm{Q}$ includes the two coefficients:
\begin{equation}
 I(\bm{Q}) = \left(\frac{1}{V} - \alpha\right) Q^0 - 4 \kappa Q^2 + \mathcal{O}(Q^4).
 \label{eq:I_Q_expansion}
\end{equation}
Noting that $\xi(\bm{k})$ and $h(\bm{k})$ are even in $\bm{k}$, and using the $d$-wave order parameter $\hat{\phi}(\bm{k}_n) = \sqrt{2} (k_{nx}^2 - k_{ny}^2) \ii\hat{\sigma}_y$, we calculate $I(\bm{Q})$ as
\begin{align}
 I(\bm{Q})
 &= T \sum_{s} \sum_{\omega_n} \int \frac{\mathrm{d}\bm{k}}{(2\pi)^2}
 \frac{2(k_{nx}^2 - k_{ny}^2)^2}{\left[\ii\omega_n - \xi(\bm{k} + \bm{Q}) + (-1)^s h(\bm{k} + \bm{Q})\right] \left[- \ii\omega_n - \xi(\bm{k} - \bm{Q}) - (-1)^s h(\bm{k} - \bm{Q})\right]} \notag \displaybreak[2] \\
 &\simeq T \sum_{s} \sum_{\omega_n} \int \frac{\mathrm{d}\bm{k}}{(2\pi)^2}
 \frac{2(k_{nx}^2 - k_{ny}^2)^2}{\left[\ii\omega_n - \xi(\bm{k}) - \bm{v}(\bm{k}) \cdot \bm{Q} + (-1)^s h(\bm{k} + \bm{Q})\right] \left[- \ii\omega_n - \xi(\bm{k}) + \bm{v}(\bm{k}) \cdot \bm{Q} - (-1)^s h(\bm{k} - \bm{Q})\right]} \notag \displaybreak[2] \\
 &\simeq \rho_{\mathrm{F}} T \sum_{s} \sum_{\omega_n} \int_{-\infty}^{\infty} \mathrm{d}\xi \Braket{
 \frac{\cos^2(2\theta_{\bm{k}})}{\left[\xi - \ii\omega_n + \bm{v}(\bm{k}) \cdot \bm{Q} - (-1)^s h(\bm{k} + \bm{Q})\right] \left[\xi + \ii\omega_n - \bm{v}(\bm{k}) \cdot \bm{Q} + (-1)^s h(\bm{k} - \bm{Q})\right]}}_{\mathrm{FS}},
 \label{eq:int_trX_tmp}
\end{align}
where $\bm{v}(\bm{k}) := \nabla_{\bm{k}} \xi(\bm{k}) = \bm{k} / m$ is the velocity.
In the last line, we used the approximation
\begin{gather}
 2 \int \frac{\mathrm{d}\bm{k}}{(2\pi)^2} (\cdot) \simeq \rho_{\mathrm{F}} \int_{-\infty}^{\infty} \mathrm{d}\xi \braket{\cdot}_{\mathrm{FS}}, \displaybreak[2] \\
 \braket{f(\bm{k})}_{\mathrm{FS}} := \int_{0}^{2\pi} \frac{\mathrm{d}\theta_{\bm{k}}}{2\pi} f(k_{\mathrm{F}} \cos\theta_{\bm{k}}, k_{\mathrm{F}} \sin\theta_{\bm{k}}),
\end{gather}
where the angular average is taken over the undistorted Fermi surface with the Fermi momentum $k_{\mathrm{F}}$.
Using the formula~\eqref{eq:formula_int_xi_1}, Eq.~\eqref{eq:int_trX_tmp} is reduced to
\begin{align}
 I(\bm{Q})
 &= \rho_{\mathrm{F}} \pi T \sum_{s} \sum_{\omega_n} \left\langle \frac{\cos^2(2\theta_{\bm{k}})}{|\omega_n| + \ii \Omega_s(\bm{k}, \bm{Q}) \sgn(\omega_n)} \right\rangle_{\mathrm{FS}} \notag \\
 &= \rho_{\mathrm{F}} \pi T \sum_{s} \sum_{\omega_n} \sum_{l=0}^{\infty} (-1)^l \left\langle \cos^2(2\theta_{\bm{k}}) \frac{\Omega_s(\bm{k}, \bm{Q})^{2l}}{|\omega_n|^{2l+1}} \right\rangle_{\mathrm{FS}} \notag \\
 &= \rho_{\mathrm{F}} \sum_{s} \sum_{l=0}^{\infty} (-1)^l S_{2l+1}(T) \left\langle \cos^2(2\theta_{\bm{k}}) \Omega_s(\bm{k}, \bm{Q})^{2l} \right\rangle_{\mathrm{FS}},
 \label{eq:int_trX}
\end{align}
where we define
\begin{align}
 \Omega_s(\bm{k}, \bm{Q})
 &:= \bm{v}(\bm{k}) \cdot \bm{Q} - \frac{(-1)^s}{2} [h(\bm{k} + \bm{Q}) + h(\bm{k} - \bm{Q})] \notag \\
 &= (-1)^{s-1} h(\bm{k}) + \bm{v}(\bm{k}) \cdot \bm{Q} + \frac{(-1)^{s-1}}{2} (\bm{Q} \cdot \nabla_{\bm{k}})^2 h(\bm{k}) + \dotsb.
\end{align}
Then we expand $\sum_{s} \Omega_s(\bm{k}, \bm{Q})^{2l}$ ($l \geq 1$) with respect to $\bm{Q}$:
\begin{equation}
 \sum_{s} \Omega_s(\bm{k}, \bm{Q})^{2l}
 = 2h(\bm{k})^{2l} + 2 \cdot \binom{2l}{1} \, h(\bm{k})^{2l-1} (\bm{Q} \cdot \nabla_{\bm{k}})^2 h(\bm{k})
 + 2 \cdot \binom{2l}{2} \, h(\bm{k})^{2l-2} [\bm{v}(\bm{k}) \cdot \bm{Q}]^2 + \dotsb.
 \label{eq:Omega_2l}
\end{equation}
From Eqs.~\eqref{eq:I_Q_expansion}, \eqref{eq:int_trX}, and \eqref{eq:Omega_2l}, therefore, the GL coefficients $\alpha$ and $\kappa$ are given by
\begin{align}
 \alpha &= \frac{1}{V} - 2\rho_{\mathrm{F}} \sum_{l=0}^{\infty} (-1)^l S_{2l+1}(T) \left\langle \cos^2(2\theta_{\bm{k}}) h(\bm{k})^{2l} \right\rangle_{\mathrm{FS}}, \\
 \kappa &= \frac{\rho_{\mathrm{F}} k_{\mathrm{F}}^2}{2m^2}
 \sum_{l=1}^{\infty} (-1)^{l-1} \binom{2l}{2} \, S_{2l+1}(T) \left\langle \cos^2(2\theta_{\bm{k}})
 \left[ \frac{2m^2 h(\bm{k})^{2l-1}}{(2l-1) k_{\mathrm{F}}^2} (\hat{\bm{Q}} \cdot \nabla_{\bm{k}})^2 h(\bm{k})
 + h(\bm{k})^{2l-2} (\bm{k}_n \cdot \hat{\bm{Q}})^2 \right] \right\rangle_{\mathrm{FS}}.
 \label{eq:GL_continuum_dS_kappa}
\end{align}
Using the formulas~\eqref{eq:formula_int_theta_1}--\eqref{eq:formula_int_theta_3}, one can reproduce Eqs.~\eqref{eq:GL_continuum_sZ_dS_alpha} and \eqref{eq:GL_continuum_sZ_dS_beta_kappa} for a uniform Zeeman field [$h(\bm{k}) = h$], and Eqs.~\eqref{eq:GL_continuum_dZ_dS_alpha} and \eqref{eq:GL_continuum_dZ_dS_kappa} for a momentum-dependent Zeeman field [$h(\bm{k}) = \sqrt{2} h (k_{nx}^2 - k_{ny}^2) = \sqrt{2} h \cos(2\theta_{\bm{k}})$].

Next, the coefficient $\beta$ is calculated in a similar way to $I(\bm{Q})$:
\begin{align}
 \beta
 &= \frac{T}{2} \sum_{\omega_n} \int \frac{\mathrm{d}\bm{k}}{(2\pi)^2}
 \tr\left[\hat{X}(\bm{k}, \bm{0}; \omega_n)^2\right] \notag \\
 &= T \sum_{s} \sum_{\omega_n} \int \frac{\mathrm{d}\bm{k}}{(2\pi)^2}
 \frac{2(k_{nx}^2 - k_{ny}^2)^4}{\left[\ii\omega_n - \xi(\bm{k}) + (-1)^s h(\bm{k})\right]^2 \left[- \ii\omega_n - \xi(\bm{k}) - (-1)^s h(\bm{k})\right]^2} \notag \\
 &\simeq \rho_{\mathrm{F}} T \sum_{s} \sum_{\omega_n} \int_{-\infty}^{\infty} \mathrm{d}\xi
 \Braket{\frac{\cos^4(2\theta_{\bm{k}})}{\left[\xi - \ii\omega_n - (-1)^s h(\bm{k})\right]^2 \left[\xi + \ii\omega_n + (-1)^s h(\bm{k})\right]^2}}_{\mathrm{FS}} \notag \\
 &= \frac{\rho_{\mathrm{F}}}{2} \pi T \sum_{s} \sum_{\omega_n} \left\langle \frac{\cos^4(2\theta_{\bm{k}})}{\left[|\omega_n| - \ii (-1)^s h(\bm{k}) \sgn(\omega_n)\right]^3} \right\rangle_{\mathrm{FS}} \notag \\
 &= \frac{\rho_{\mathrm{F}}}{2} \pi T \sum_{s} \sum_{\omega_n} \sum_{l=1}^{\infty} (-1)^{l-1} \binom{2l}{2} \left\langle \cos^4(2\theta_{\bm{k}}) \frac{h(\bm{k})^{2l-2}}{|\omega_n|^{2l+1}} \right\rangle_{\mathrm{FS}} \notag \\
 &= \rho_{\mathrm{F}} \sum_{l=1}^{\infty} (-1)^{l-1} \binom{2l}{2} \, S_{2l+1}(T) \left\langle \cos^4(2\theta_{\bm{k}}) h(\bm{k})^{2l-2} \right\rangle_{\mathrm{FS}},
\end{align}
\end{widetext}
where we used the formula~\eqref{eq:formula_int_xi_2}.
With the help of the formula~\eqref{eq:formula_int_theta_1}, this equation results in Eq.~\eqref{eq:GL_continuum_sZ_dS_beta_kappa} for a uniform Zeeman field, and Eq.~\eqref{eq:GL_continuum_dZ_dS_beta} for a $d$-wave Zeeman field.

\subsection{The case of strongly coupled superconductivity}
\label{app:GL_coefficients_continuum_strong}
We make a short notice on the GL coefficient $\kappa$ in a $d$-wave Zeeman field.
Equation~\eqref{eq:GL_continuum_dZ_dS_kappa} explicitly contains the Fermi energy $E_{\mathrm{F}}$, which arises from the derivative of the momentum-dependent field [the first term in Eq.~\eqref{eq:GL_continuum_dS_kappa}].
In the main text, we neglected this term since weakly coupled superconductors satisfy $T_{\mathrm{c}0} \ll E_{\mathrm{F}}$.

\begin{figure}[b]
 \includegraphics[width=.7\linewidth, pagebox=artbox]{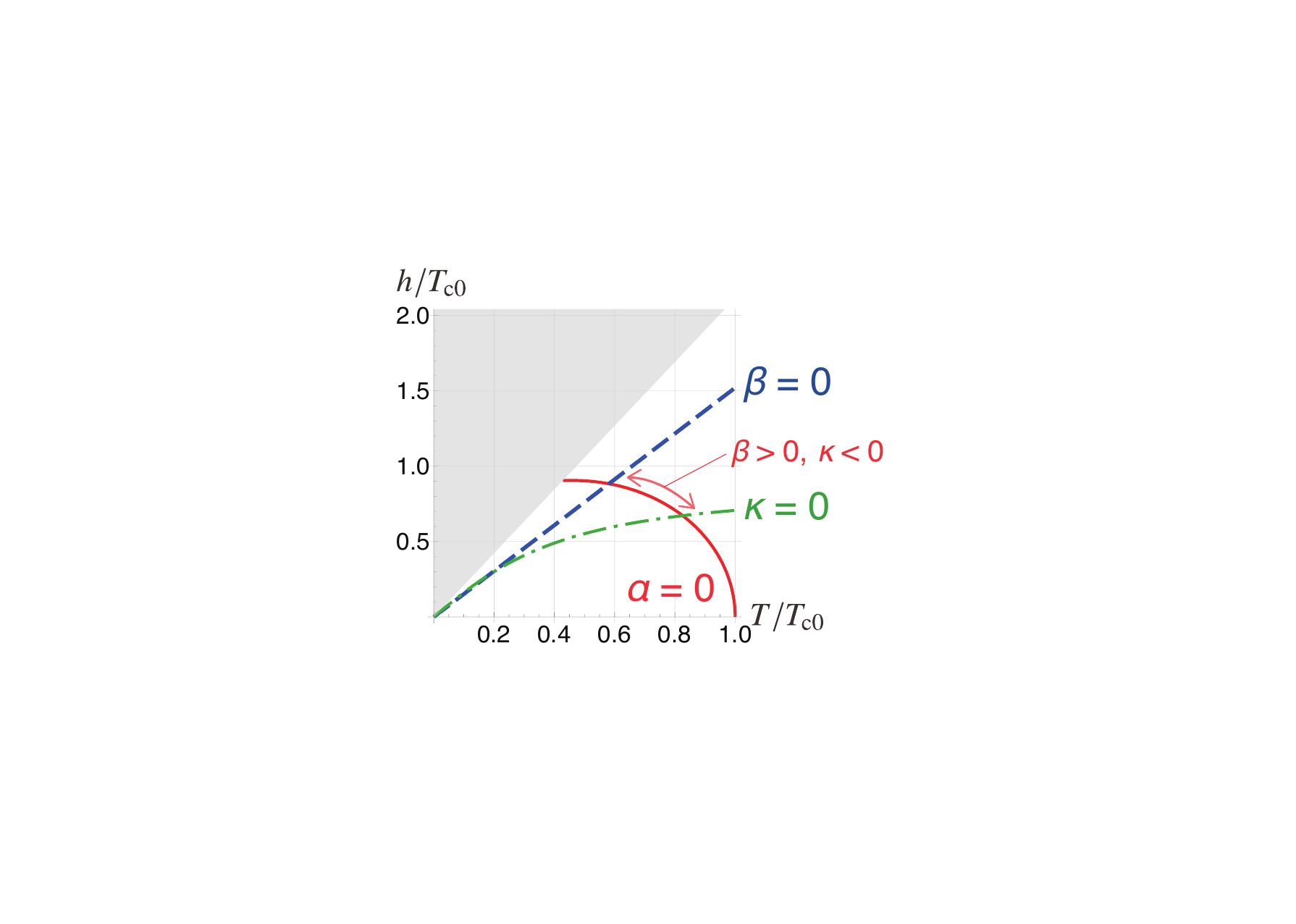}
 \caption{$T$--$h$ phase diagram in the continuum model with a $d$-wave Zeeman field, for strongly coupled superconductivity. As in Fig.~\ref{fig:phase_GL_continuum_dS}, the $\alpha = 0$ lines are unreliable when $h / T$ is large (dark shaded areas) since we truncate the infinite series after 100 terms.}
 \label{fig:phase_GL_continuum_dZ_dS_strong}
\end{figure}

Here, we consider strongly coupled superconductors where the transition temperature $T_{\mathrm{c}0}$ is comparable to the Fermi energy $T_{\mathrm{c}0}$.
Figure~\ref{fig:phase_GL_continuum_dZ_dS_strong} show a phase diagram for $E_{\mathrm{F}} / T_{\mathrm{c}0} = 1$.
In this case, the zeros of $\kappa$ (green dash-dotted line) become a convex curve upward, while they form a straight line in the weak-coupling limit [Fig.~\ref{fig:phase_GL_continuum_dS}(b)].
Therefore, the FF state is stabilized in the intermediate field regime between the blue and green lines.
In a higher Zeeman field, however, the coefficient $\beta$ also flips to negative; then the amplitude-modulated superconductivity becomes more stable.
Although this behavior does not describe our results in Fig.~\ref{fig:phase_GL_square2}, this may correspond to a previous theoretical analysis reporting the transition from the FF state to the PDW state by changing the chemical potential~\cite{Sim2024}.

\subsection{Integral formulas}
\label{app:GL_coefficients_continuum_formulas}
We here introduce some integral formulas that are used in the analytical calculations above.
First, we show integrals with respect to the energy $\xi$.
Let $a$ and $b$ be arbitrary real numbers.
Then the following equations are satisfied:
\begin{align}
 &\int_{-\infty}^{\infty} \mathrm{d}\xi \frac{1}{[\xi - (\ii\omega_n + a)] [\xi - (-\ii\omega_n + b)]} \notag \\
 &= \frac{\pi}{|\omega_n| + \ii \frac{b-a}{2} \sgn(\omega_n)},
 \label{eq:formula_int_xi_1} \\
 & \int_{-\infty}^{\infty} \mathrm{d}\xi \frac{1}{[\xi - (\ii\omega_n + a)]^2 [\xi - (-\ii\omega_n + b)]^2} \notag \\
 &= \frac{\pi / 2}{\left[|\omega_n| + \ii \frac{b-a}{2} \sgn(\omega_n)\right]^3},
 \label{eq:formula_int_xi_2}
\end{align}
which are easily proved by using the residue theorem.

\begin{widetext}
Next, we introduce integrals with respect to the angle $\theta_{\bm{k}}$.
Let $l$ be an integer greater than or equal to $1$.
Then
\begin{gather}
 \int_{0}^{2\pi} \frac{\mathrm{d}\theta_{\bm{k}}}{2\pi} \cos^{2l}(2\theta_{\bm{k}})
 = \frac{(2l)!}{2^{2l} (l!)^2},
 \label{eq:formula_int_theta_1} \\
 \int_{0}^{2\pi} \frac{\mathrm{d}\theta_{\bm{k}}}{2\pi} \cos^{2l}(2\theta_{\bm{k}}) \cos^2\theta_{\bm{k}}
 = \int_{0}^{2\pi} \frac{\mathrm{d}\theta_{\bm{k}}}{2\pi} \cos^{2l}(2\theta_{\bm{k}}) \sin^2\theta_{\bm{k}}
 = \frac{(2l)!}{2^{2l+1} (l!)^2},
 \label{eq:formula_int_theta_2} \\
 \int_{0}^{2\pi} \frac{\mathrm{d}\theta_{\bm{k}}}{2\pi} \cos^{2l-1}(2\theta_{\bm{k}}) \sin^2\theta_{\bm{k}} (\sin^2\theta_{\bm{k}} - 3\cos^2\theta_{\bm{k}})
 = \int_{0}^{2\pi} \frac{\mathrm{d}\theta_{\bm{k}}}{2\pi} \cos^{2l-1}(2\theta_{\bm{k}}) \cos^2\theta_{\bm{k}} (3\sin^2\theta_{\bm{k}} - \cos^2\theta_{\bm{k}})
 = - \frac{(2l)!}{2^{2l+1} (l!)^2},
 \label{eq:formula_int_theta_3}
\end{gather}
are satisfied.
\end{widetext}


%

\end{document}